\documentclass[prb,twocolumn]{revtex4-1}
\usepackage{float}
\usepackage[caption = false]{subfig}
\usepackage[final]{graphicx}

\usepackage{dcolumn}
\usepackage{bm}
\usepackage{amssymb,amsmath}
\usepackage{color}
\usepackage{epstopdf}

\begin{document}
	\title{Correlated disorder in a well relaxed model binary glass through a local SU(2) bonding topology}
	\author{P. M. Derlet}
	\email{Peter.Derlet@psi.ch} 
	\affiliation{Condensed Matter Theory Group, Paul Scherrer Institut, CH-5232 Villigen PSI, Switzerland}
	\date{\today}

\begin{abstract}
A quantitative understanding of the microscopic constraints which underlie a well relaxed glassy structure is the key to developing a microscopic theory of structural evolution and plasticity for the amorphous solid. Here we demonstrate the applicability of one such theory of local bonding constraints developed by D. R. Nelson [Phys. Rev. B 28, 5515 (1983)], for a model binary Lennard-Jones glass structure that has undergone an isothermal annealing simulation spanning over 10 micro-seconds of physical simulation time. By introducing a modified radical Voronoi tessellation which removes some ambiguity in how nearest neighbour bonds are enumerated, it is found, that a large proportion ($>95\%$) of local atomic environments follow the connectivity rules of the SU(2) topology of Nelson's work resulting in a dense network of disclination lines characterizing the defect bonds. Furthermore, it is numerically shown that a low energy glass structure corresponds to a reduced level of bond-length frustration and thus a minimally defected bond-defect network. It is then demonstrated that such a defect network provides a framework in which to analyse thermally-activated structural excitations, revealing those high-energy/low-density regions not following the connectivity constraints are more likely to undergo structural rearrangement that often results in a local relaxation that ends with the creation of new SU(2) local topology content.
\end{abstract}

\maketitle

\section{Introduction}

If the temperature of a glass forming metallic liquid is lowered sufficiently fast below the melting point, the nucleation and growth processes needed for crystallization do not occur and the material enters the meta-stable under-cooled liquid phase. Kauzmann, in his seminal paper of 1948~\cite{Kauzmann1948}, asserted that as the temperature continues to drop, fluctuation processes rare to the equilibrium liquid become increasingly likely. These microscopic fluctuations, broadly known as dynamical heterogenities~\cite{Sillescu1999,Ediger2012}, have characteristic timescales and length-scales which rapidly increase with decreasing temperature and when these become larger than the relevant observational timescales, the material falls out of the meta-equilibrium of the under-cooled liquid and enters that of the bulk-metallic glass~\cite{Kauzmann1948,Goldstein1969,Debenedetti2001}.

The materials science approach is to quantify the nature of this glassy meta-equilibrium regime --- its structural characteristics through its fluctuations and instabilities, and ultimately their collective and macroscopic response to stimuli such as an external heating and/or loading protocols~\cite{Chaudhari1978,Hufnagel2016,Sun2016}. With no long-range order, understanding amorphous structure has to be based on quantifying the atomic-scale constraints. Indeed, the configurational entropies of a glass can be up to an order of magnitude lower than its liquid phase at the melting temperature resulting in a considerable reduction of the accessible structural phase space~\cite{Debenedetti2001}. The classification of such constraints, and how they may be broken, can lead to the notion of a structural defect hierarchy which in turn could be the basis for a microscopic theory of the amorphous solid. 

Kauzmann goes on to assert that deep within the under-cooled liquid those microscopic structural fluctuations necessary for crystal nucleation, and those deemed to be rare in the equilibrium liquid state, have comparable time-scales, competing with each other to determine the meta-equilibrium structure of the under-cooled liquid, and therefore that of the resulting glass structure. This suggests that the sought after atomic-scale constraints of the glass feature aspects of local structural motives compatible with local crystalline order~\cite{Derlet2020,An2020}.

\begin{figure*}
	\includegraphics[width=0.65\linewidth,trim=3cm 2cm 3cm 2cm,clip]{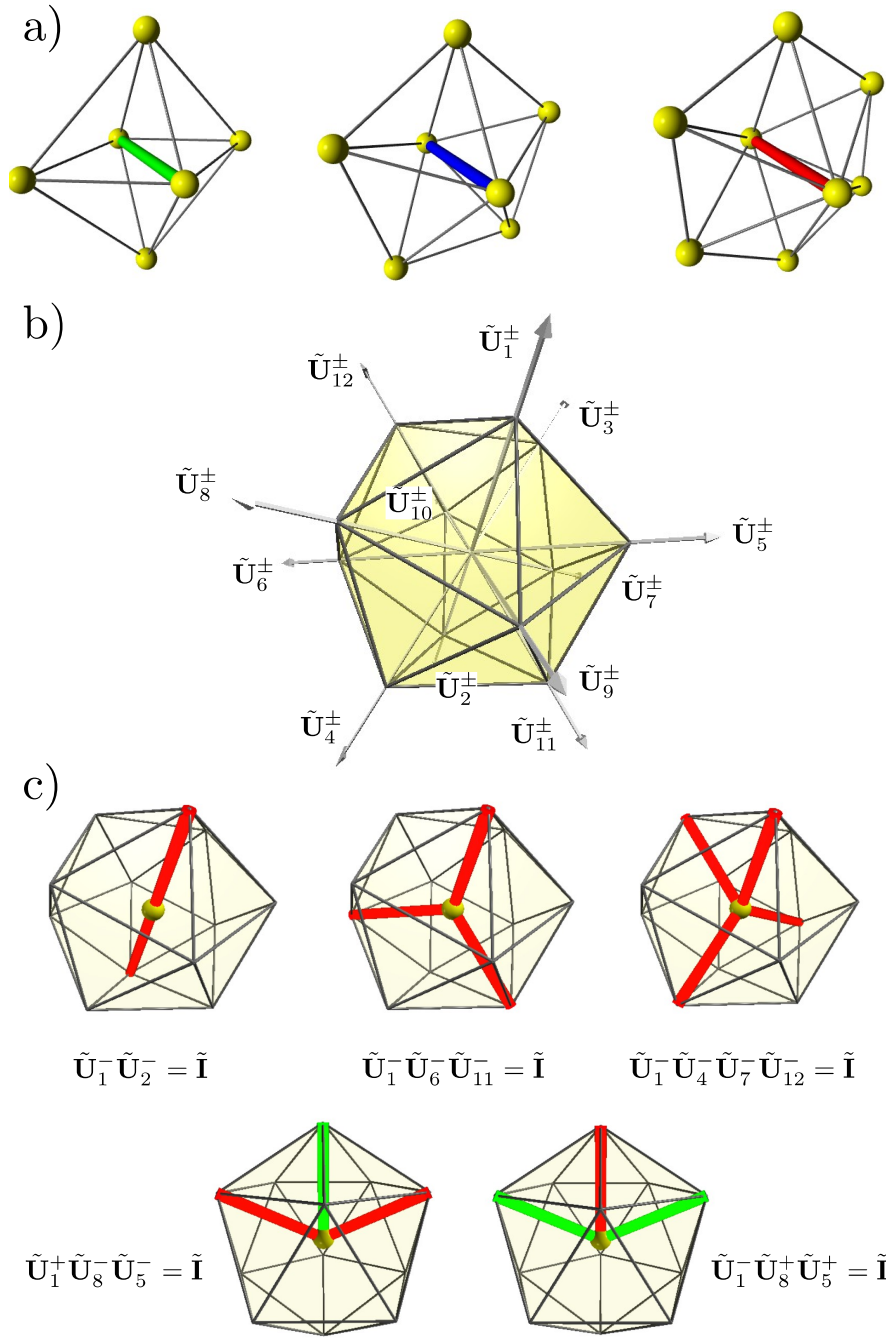}
	\caption{a) Four (green), five (blue) and six (red) fold bonds, defined by the number of common neighbours or the number of tetrahedra that may be packed around the bond. b) Yellow shaded icosahedron whose five-fold rotation symmetry axis are labeled from $i=1$ to 12. To each such axis corresponds the operator, $\tilde{\mathbf{U}}^{\pm}_{i}$, which creates a 4-fold ($+$) or a 6-fold ($-$) bond defect. c) Examples of allowed bond defects where green/red defect bonds represent 4-fold/6-fold bonds. Here the upper panels represent examples of Frank-Kasper polyhedra~\cite{Frank1958} and the lower panels examples of Nelson polyhedra~\cite{Nelson1983a,Nelson1983b}.}
	\label{FigGeneral}
\end{figure*}

Insight into those structural fluctuations that are not compatible with general crystalline order, and which are rare in the liquid, can be gained by considering the minimum energy configuration of a quadruple of atoms~\cite{Chaudhari1978}. For a mono-atomic system this configuration results in a regular tetrahedron. However, such regular tetrahedra can not be packed in a volume filling way. The packing of twelve distorted tetrahedrons around a single atom, forms an icosahedron, and represents the minimally distorted tetrahedral packing with the bond to the central atom being shorter than that between its neighbours. The common neighbours between the central atom and one of its neighbours, number five, and represent a packing of five tetrahedra. This is referred to as a 5-fold bond (Fig.~\ref{FigGeneral}a). Continuing a tetrahedral packing protocol, whilst maintaining the icosahedral point symmetry soon results in distorted tetrahedra and strong internal strain, that can be alleviated by breaking the icosahedral point symmetry and introducing ``defects'' in the form of 4-fold and 6-fold bonds --- an idea proposed by Frank and Kasper~\cite{Frank1958} where generally an $n$-fold bond is represented by the packing of $n$ distorted tetrahedra around the bond axis (Fig.~\ref{FigGeneral}a). It is a majority of these 4-fold and 6-fold bonds that lead to long range crystalline order, resulting in these bonds being referred to as ``crystal-like'' bonds, whereas the 5-fold bond is referred to as the ``liquid-like'' bond (which, in fact, is rare in the equilibrium liquid). Relative to the Z12 coordinated local structure of the icosahedral environment, Frank and Kasper~\cite{Frank1958} identified these bonds as higher coordinated environments involving two, three, four additional 6-fold bonds whose coordinations are respectively 13, 14 and 15, whereas Bernal~\cite{Bernal1964} identified lower coordinated environments as involving 2, 3 and 4-fold bonds whose coordinations are respectively Z11, Z10 and Z9.

The discussed structural motives are all relevant for the amorphous structure, as evidenced by the success of the glass structural models of Miracle and co-workers~\cite{Miracle2003,Miracle2004,Miracle2006,Laws2015}, and the use of icosahedral content and the polytope variants pioneered by Ma and co-workers~\cite{Sheng2006,Ding2014,Ma2015}. The work of Miracle demonstrates that structural insight into binary and ternary metallic glasses can be gained through an analysis of dense local packings of solute atoms which take into account the coordination of the neighbouring atoms~\cite{Miracle2003}, whilst the work of Ma demonstrates the central importance of local icoshadral environments as a structural measure of relaxation and rejuvenation in bulk metallic glasses. Indeed, J\'{o}nsson and Andersen~\cite{Jonsson1988} had showed somewhat earlier, that as the glass-transition temperature regime is entered and passed, a local icosahedral ordering rapidly percolates throught out the model Lennard Jones amorphous solid. Together, these and many subsequent works introduce the notion of medium range order, through the connectivity of how such local structural units may be packed. 

Atomistic simulation of well relaxed model binary alloys has given some insight into the nature of such medium range order. The work of Zemp {\em et al}~\cite{Zemp2014,Zemp2016} has demonstrated for a model CuZu glass, that high temperature annealing simulations spanning up to 800 nsec result in atomic environments whose icosahedrally coordinated atoms form local bond networks identified as fragments of the C15 Laves phase --- a crystal structure belonging to the class of close-packed topological alloy structures which are dominated by 4-fold and 6-fold bonds. This has also been observed using the Wahnstr\"{o}m~\cite{Wahnstrom1991} Lennard Jones potential~\cite{Pedersen2010,Derlet2020} --- a potential that is known to nucleate Laves crystals at high enough temperatures~\cite{Pedersen2007}. 

The notion of 4-fold and 6-fold bonds being defected bonds was put on a firm theoretical founding by Nelson~\cite{Nelson1983a,Nelson1983b}. Central to this approach was the focus on the geometrical frustration associated with the inability to simultaneously minimize the energy of all bonds in three spatial dimensions~\cite{Chaudhari1978}. Here, the 5-fold bond and its packing of five distorted tetraherons represents a locally minimally frustrated structure at the individual bond-length scale, whereas the icosahedral environment of twelve 5-fold bonds represents the minimally frustrated structure of the nearest neighbour environment length-scale. Inspired by homotopy theory~\cite{Mermin1979}, Nelson demonstrated that 4-fold and 6-bold bond defects at this latter length scale are restricted to a subset of nearest neighbour environments --- see Sec.~\ref{ssec:su2}. The connectivity of these restricted local topologies could then represent a robust mathematical representation of the amorphous structure, defining the glassy structure in terms of microscopic constraints that the local atomic environments must satisfy in which any low energy microscopic realization would be described by a network of $n(\ne5)$-fold bond defect lines --- the so-called 4- and 5-fold disclination line defects. 

By performing molecular dynamics on a 500 atom model Mg$_{3}$Ca$_{7}$ system, Qi and Wang~\cite{Qi1991} found several realizations of the local topologies predicted by Nelson, confirming that such local structural motives do exist. Because of their rarity, presumably due to the glassy structures being relatively unrelaxed, an extended connected array of disclination line defects was not demonstrated. It is the goal of the present work to demonstrate that a well-known model binary glass system, when well relaxed, does indeed exhibit such a network of disclinations involving almost all parts of the amorphous structure. 

The paper is organised as follows. In Sec.~\ref{sec:methods} the relevant aspects of Nelson's work is introduced leading to a mathematical theory of local constraints based on an SU(2) algebra, the atomistic model binary glass simulations are described to which the SU(2) algebra is applied, and a modified radical Voronoi tessellation is developed that is better able to define the local nearest neighbour geometry necessary to define the $n$-fold bond defect structure of the produced microscopic samples. The results section, Sec.~\ref{sec:results},  details the isothermal atomistic simulations which involve approximately 14 billion molecular dynamics iterations ($\sim14$ micro-seconds of physical simulation time) facilitating significant structural relaxation (Sec.~\ref{ssec:md}), the application of the SU(2) algebra to these glassy structures (Secs.~\ref{ssec:nfold} and \ref{ssec:su2app}), and how this description correlates with traditional atomistic structural descriptors such as local atomic volume (free volume), local energy, local stress, frustration, and spatial structural correlation. These latter aspects provide strong numerical evidence that a minimally SU(2) defected glassy structure does indeed correspond to a structure with a reduced bond frustration. Sec.~\ref{sec:dis} discusses how the present work relates to the efficient packing glass structure model of Miracle and co-workers~\cite{Miracle2003,Miracle2004,Miracle2006,Laws2015} and how the SU(2) framework allows for an entirely new aspect to studying thermally-activated localized structural excitations and how they can lead to structural relaxation with the creation of minimally defected content.

\section{Theoretical and simulation methodology} \label{sec:methods}

\subsection{The SU(2) algebra of local structure} \label{ssec:su2}

Nelson begins with the flat defect-free 2D triangular lattice whose atoms are Z'5 coordinated. The Z' terminology refers to nearest neighbour coordinations in 2D. The associated Z'4 and Z'6 local coordination defect environments are topological (referred to as disclinations) and maybe associated with the homotopy group arising from the translational and rotational symmetry of the liquid phase in two dimensions, and the discrete lattice symmetry of the solid 2D triangular phase. This 2D picture is then applied to the surface of a 3-sphere where the discrete lattice symmetry is replaced with the discrete symmetries of the icosahedron --- the curved and finite 2D triangular lattice analogue. The defect free atomic structure would represent the environment of the Z12 coordinated icosahedron structure of an atom within a 3D glass, and the Z'4 and Z'6 defects are now the 4-fold and 6-fold bonds referred to earlier. See Fig.~\ref{FigGeneral}. 

In what follows only the 5-fold symmetry of the icosahedron are considered, and in particular the conjugacy classes associated with $\pm72$ rotations around the 6+6 five-fold axes (See Sec.~IIC in Ref.~\cite{Nelson1983b}) which may be identified as the fundamental disclinations of the icosahedron solid. The corresponding group elements of these defect types forms an SU(2) algebra, defined by the matrices:
\begin{equation}
\tilde{\mathbf{U}}^{\pm}_{i}=\exp\left(\frac{1}{2}i\omega^{\pm} \mathbf{n}_{i}.\tilde{\boldsymbol{\sigma}}\right), \label{eqnsu2}
\end{equation}
where $\mathbf{n}_{i}$ are the twelve 5-fold rotation axes of the icosahedron (see Fig.~\ref{FigGeneral}b), $\omega^{+}=72^{\circ}$ and $\omega^{-}=-72^{\circ}$ refer to the so-called disclinations associated with, respectively, the 4- and 6-fold bonds, and $\boldsymbol{\sigma}$ is the 3-vector of Pauli matrices. The allowed $n$-fold bond structure of a particular atom is then defined by products of the $\tilde{\mathbf{U}}^{\pm}_{i}$ giving the identity matrix. For example the Frank-Kasper Z14 polyhedron~\cite{Frank1958}, could be given by $\tilde{\mathbf{U}}^{-}_{1}\tilde{\mathbf{U}}^{-}_{2}=\tilde{\mathbf{I}}$, the Frank-Kasper Z15 polyhedron by $\tilde{\mathbf{U}}^{-}_{1}\tilde{\mathbf{U}}^{-}_{5}\tilde{\mathbf{U}}^{-}_{11}=\tilde{\mathbf{I}}$, and the Frank-Kasper Z16 polyhedron by $\tilde{\mathbf{U}}^{-}_{1}\tilde{\mathbf{U}}^{-}_{8}\tilde{\mathbf{U}}^{-}_{9}\tilde{\mathbf{U}}^{-}_{12}=\tilde{\mathbf{I}}$ matrix product. The Bernal-hole bond order structures~\cite{Bernal1964} are given by similar products, but in terms of the $\mathbf{U}^{+}_{i}$. Ref.\cite{Nelson1983b} also introduces the so-called vacancy (free-volume defect) in which one 6-fold bond is created and two 4-bonds are created, as well as interstitial (negative free-volume defect) in which one 4-fold bond and two 6-fold bonds are created at a site. These polyhedra will be referred to as Z11 and Z13 Nelson polyhedra. Algebraically these take the respective forms of  $\tilde{\mathbf{U}}^{-}_{1}\tilde{\mathbf{U}}^{+}_{5}\tilde{\mathbf{U}}^{-}_{12}=\tilde{\mathbf{I}}$ and $\tilde{\mathbf{U}}^{+}_{1}\tilde{\mathbf{U}}^{+}_{6}\tilde{\mathbf{U}}^{-}_{9}=\tilde{\mathbf{I}}$. The Frank-Kasper and Nelson polyhedra are visualized in terms of the geometry entailed by eqn.~\ref{eqnsu2} in Fig.~\ref{FigGeneral}c. There are many more allowed defected bond order structures, some of which will be studied in more detail in what follows. 

The implication for connectivity of defected bond-order network satisfying this algebra throughout the amorphous structure is clear. A network of disclination lines whose inter-connectivity satisfies the local need for the product of the correspond matrices to give the identity operator. As pointed out by Nelson, an immediate consequence is that a 4-fold or 6-fold bond, disclination line, cannot terminate at a site that is otherwise defect free since obviously, $\tilde{\mathbf{U}}^{\pm}_{i}\ne\tilde{\mathbf{I}}$.

\subsection{Model binary glass sample preparation}

For the present work, the model binary Lennard-Jones potential of Wahnstr\"{o}m~\cite{Wahnstrom1991} is used. This potential is able to capture the essential structural physics of bulk binary metallic glasses such as that of CuZr~\cite{Frank1952,Chaudhari1978,Sheng2006,Ma2015}. The Lennard-Jones potential has the following form:
\begin{equation}
V_{ab}(r)=4\varepsilon\left(\left(\frac{r}{\sigma_{ab}}\right)^{12}-\left(\frac{r}{\sigma_{ab}}\right)^{6}\right) \label{EqnLJ}
\end{equation}
where $\sigma_{22}=5/6\sigma_{11}$ and  $\sigma_{12}=\sigma_{21}=11/12\sigma_{11}$. Type 1 atoms are considered as the larger atom type. The atomic masses of the two atom types are arbitrarily chosen such that $m_{1}/m_{2}=2$. For a molecular dynamics iteration, a time step of 0:002778$\sigma_{11}\sqrt{m_{1}/\varepsilon}$ is used. The distance unit is taken as $\sigma=\sigma_{11}$ and the energy unit as $\varepsilon$, with stresses in the units of $\varepsilon/\sigma^{3}$. Absolute temperature is expressed as an energy $k_{\mathrm{B}}T$ with $k_{\mathrm{B}}=8.617\times10^{-5}\varepsilon$. For this work, the potential is truncated to a distance 2.5$\sigma$. All molecular dynamics (MD) simulations are performed using the LAMMPS software platform~\cite{Plimpton1995} and atomistic visualization is performed using OVITO~\cite{Stukowski2010}.

Presently, a 32000 atom 50:50 chemical concentration of small to large atoms is considered, which is the potential's eutectic composition. Periodic boundary conditions are used. Sample preparation is similar to that used in past work and involves the following steps within a fixed volume ensemble: a) a well-equilibriated high temperature liquid is prepared at a volume per atom equal to 1.3078 $\sigma^{3}$ and temperature equal to $k_{\mathrm{B}}T=1.7324\varepsilon$; b) this configuration is quenched to an intermediate high temperature of $k_{\mathrm{B}}T=0.8617\varepsilon$ until all thermodynamic quantities fluctuate around their mean values; c) after which the temperature is decreased to a value close to zero at the rate $k_{\mathrm{B}}T=0.8617\time10^{-6}\varepsilon$ per MD step; d) the fictive glass transition temperature is then defined ($T^{NVT}_{\mathrm{f}}$) as the intersection of the low temperature (amorphous solid) and high temperature (under-cooled liquid) linear temperature dependencies of the internal energy and is equal to $k_{\mathrm{B}}T^{NVT}_{\mathrm{f}}\approx0.55\varepsilon$; e) fixed volume iso-thermal simulations are then performed for approximately 14 billion MD steps, starting from the configuration at a temperature closest to 0.95$T^{NVT}_{\mathrm{f}}$ (obtained from the initial linear quench). When $\varepsilon$ is taken as 1 electron Volt, $\sigma$ as one Angstrom and the atomic masses typically as that of a metallic atom, one MD iteration corresponds to approximately a femto-second and therefore one billion MD iterations corresponds to approximately one micro-second.

Past  work~\cite{Swayamjyoti2014,Swayamjyoti2016,Derlet2017a,Derlet2018,Derlet2020} has used such NVT samples to allow for a simplified structural analysis as a function of the degree of relaxation which does not involve a global volume relaxation. Global stress relaxation does however occur, in which the initial positive global pressure switches to a negative global pressure indicating that samples would contract under fixed zero pressure conditions~\cite{Derlet2017}. Subsequent work has found that the same structural state can be obtained using either NVT or NPT ensembles, or a combination of both~\cite{Derlet2018}.

\subsection{Modification of Voronoi tesselation} \label{ssec:voro}

To determine the population of bonds within a given structure, a variant of the Voronoi tesselation~\cite{Voronoi1907,Voronoi1908,Delaunay1934} is used for systems with differently sized atoms. The radical Voronoi tesselation~\cite{Gellatly1982} (RV) gives the nearest neighbours of each atom in a poly-dispersed system, which can be then used to determine the common nearest neighbours of two neighbouring atoms. It is assumed that such a radical Voronoi tessellation takes into account all relevant atom type dependencies. This is a working assumption. From this, the $n$-fold nature of each bond is determined via the number of common nearest neighbours the two atoms, forming the bond, have. This approach is motivated by the common neighbour method of Honeycutt and Anderson~\cite{Honeycutt1987}. It is noted that the number of edges the corresponding Voronoi face has, has also been used to define the $n$-fold bond nature. However, the obtained value is generally not equal to the explicit common neighbour approach, tending to overestimate the value of $n$ due to edge contributions of more distant (non-nearest neighbour) atoms. 

\begin{figure}
	\includegraphics[width=0.95\linewidth,trim=1cm 5cm 1cm 5cm,clip]{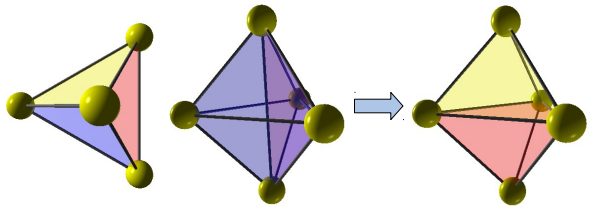}	
	\caption{The standard Voronoi tessellation can produce nearest neighbour bonds for which three common neigbours are themselves neighbours. Upon tetrahedral tessellation along such identified bonds, three strongly distorted tetrahedra are generated. A more efficient tetrahedral tessellation can be chosen by no longer identifying the original pair of atoms as a nearest neighbour bond, resulting in a tetetrahedral tessellation involving two less distorted tetrahedrons. The two left-most panels represent the original Voronoi tessellation consisting of three strongly distorted tetrahedra, whereas the right most shows the resulting two tetrahedron tessellation.}
	\label{FigModRV}
\end{figure}

The above common neighbour approach to determining the $n$-fold nature of a bond  also becomes problematic for certain common neighbour structures. These local atomic arrangements are characterized by three of the common neighbours being themselves neighbours, and whose triangular face intersects the line defined by the bond. This leads to an ambiguity in the choice of $n$. Should the remaining $n-3$ common neighbours be ignored and the bond relabeled as 3-fold, or should it be considered as a strongly distorted $n$-fold bond, or should the bond itself not be considered a nearest neighbour bond? The latter approach is presently taken, and all such bonds are removed from the list of identified nearest-neighbours. This is further motivated by the observation that a tetrahedral tessellation of such a three-fold bond involves three strongly distorted tetrahedra, a configuration which may be always replaced by two less distorted tetrahedra (see Fig.~\ref{FigModRV}). Such a modification changes the populations of bond types, removing all 3-fold bonds and many greater than 6-fold bonds, whilst increasing the 4-, 5- and 6-fold bond populations. This course of action is also motivated by the observation that the 3-fold bond originating from the homotopy analysis~\cite{Nelson1983b} may be represented by products of four-fold and six-fold bonds. 

Sec.~\ref{ssec:nfold} will show that such a modification to the Voronoi construction, referred to as the modified radical Voronoi (modRV) tessellation, results in a glassy structural classification that is better able to satisfy the Euler relation between the number of faces, edges and vertices and therefore the $n$-fold bond connectivity described by the SU(2) formalism of sec.~\ref{ssec:su2}. 

\section{Results} \label{sec:results}

\begin{figure*}
	\begin{center}
\subfloat[]{\includegraphics[width=0.4\linewidth,trim=1.5cm 0.5cm 1.5cm 1.5cm]{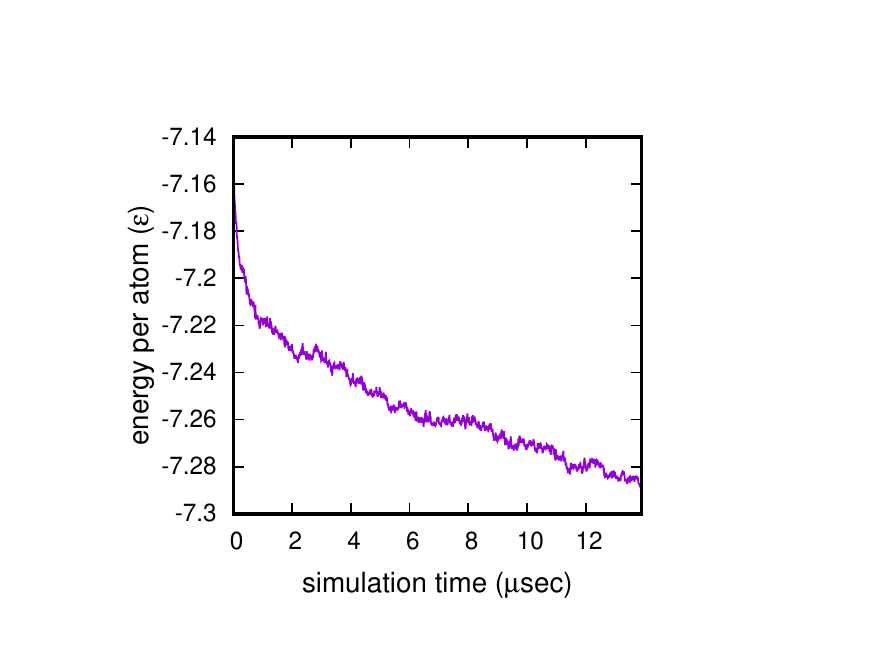}} 
\subfloat[]{\includegraphics[width=0.4\linewidth,trim=1.5cm 0.5cm 1.5cm 1.5cm]{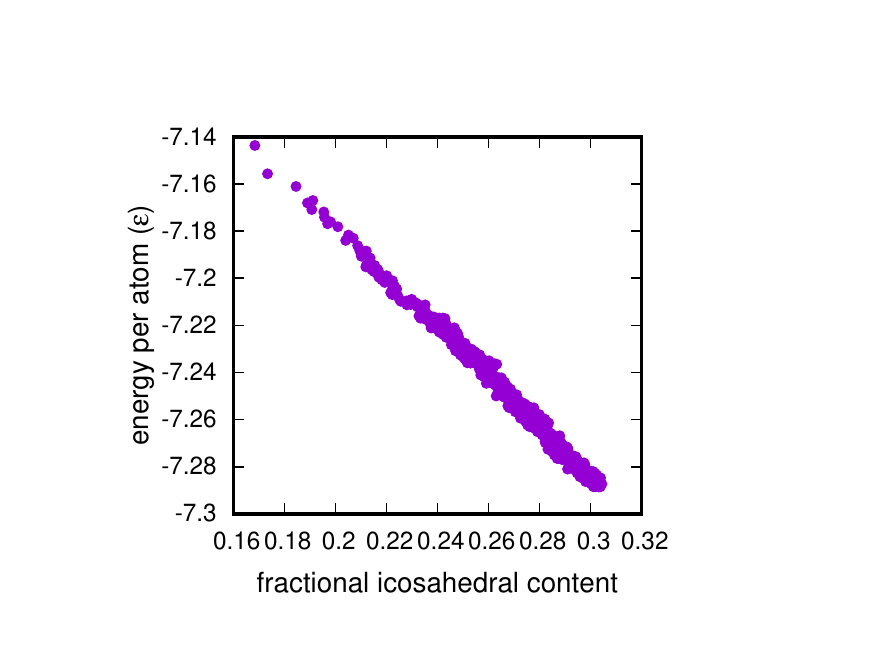}} \\		
\subfloat[]{\includegraphics[width=0.4\linewidth,trim=1.5cm 0.5cm 1.5cm 1.5cm]{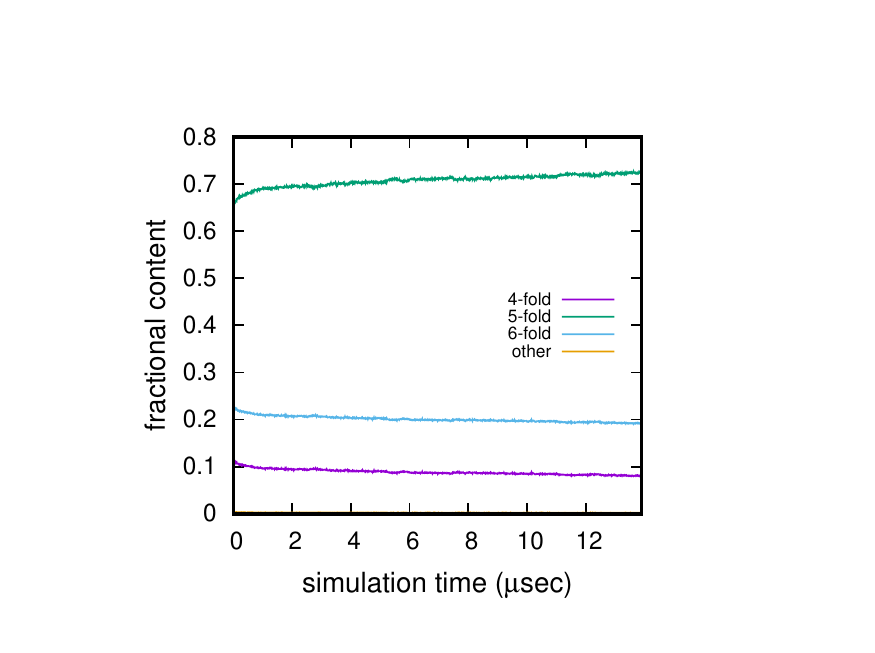}} 
\subfloat[]{\includegraphics[width=0.4\linewidth,trim=1.5cm 0.5cm 1.5cm 1.5cm]{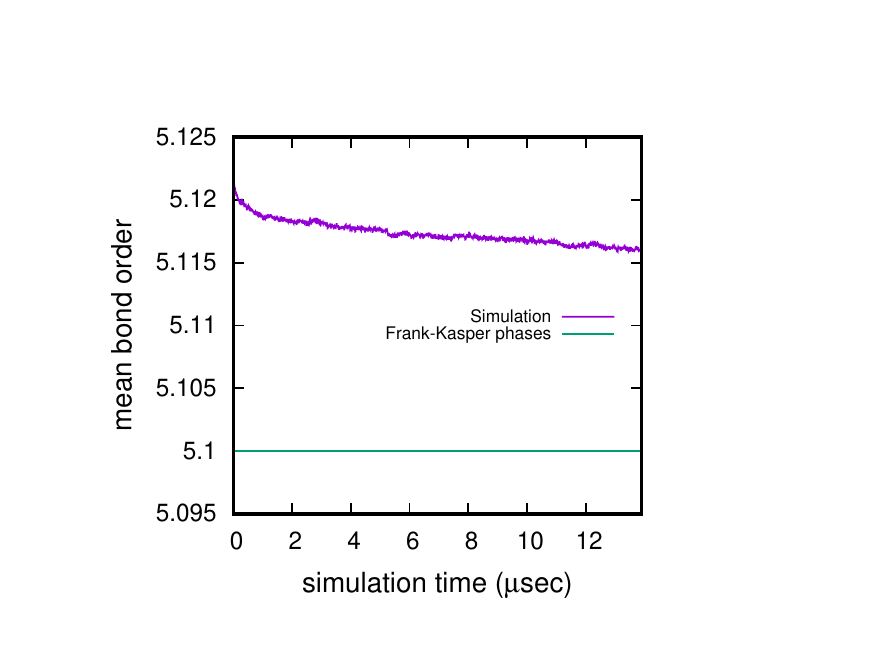}} \\
	\end{center}
	\caption{Iso-thermal glassy structural evolution during a time period of approximately 14 micro-seconds: a) plot average potential energy per atom as a function of time, b) Scatter plot of cohesive energy per atom versus fractional icosahedral content, and evolution of c) 4-fold, 5-fold, 6-fold and ``other'' fractional bond populations, and d) mean bond order. In c) the ``other'' bond population is $\sim0.3\%$, and in d) the mean bond order for the Frank-Kasper phases.}
	\label{FigEvolution}
\end{figure*}

\subsection{Structural relaxation over a timescale of order 10 micro-seconds} \label{ssec:md}

All structural analysis is performed on atomic configurations that have been quenched from their finite temperature configurations using the conjugate gradient method. This removes thermal potential energy and thus atomic displacements arising from thermal motion, placing the configuration at a local potential energy minimum --- a so-called inherent state~\cite{Stillinger1982,Stillinger1986}.

Fig.~\ref{FigEvolution}a displays the average potential energy per atom evolution during the high-temperature annealing simulations, showing a gradual drop in energy indicating structural relaxation. The underlying atomic scale processes which mediate this relaxation have been studied in detail in Refs.~\cite{Derlet2017,Derlet2020}. These configurations are extremely relaxed samples, with most published work considering glassy structures arising from sub-micro-second annealing procedures. Using the modified radical Voronoi tessellation outlined in Sec.~\ref{ssec:voro} the icosahedral content is determined as a function of time and represented as a scatter plot with respect to the average potential energy per atom in Fig.~\ref{FigEvolution}b. The data shows a strongly linear relationship compatible with other samples of past work~\cite{Derlet2017,Derlet2020}. Such a relation has also been observed for material specific embedded atom model binary potentials~\cite{Ding2014}, and is also followed for structural relaxation occurring at temperatures well below the glass transition regime, indicating that it represents a quite general relaxation trajectory that possibly ends with the formation of a nano-phase structure of C15 laves crystallites within an amorphous matrix consisting of primarily the larger atom~\cite{Derlet2020}. Ref.~\cite{Derlet2020} found that such a linear relation with icosahedral content also extends to pressure for NVT simulations and volume for NPT simulations demonstrating that the creation of icosahedral environments not only lowers the energy, but also increases the density of the structure. Indeed, the overall pressure evolution over the 14 microsecond isotherm switches sign at approximately 5 microseconds leading to a structure that would contract when using an NPT ensemble at fixed zero pressure.

\subsection{Classification of local $n$-fold bond structure} \label{ssec:nfold}

For each atom, the notation $(N_{4},N_{5},N_{6})$ is now used. Here $N_{n}$ is the number of $n$-fold bonds. The coordination $Z$ is then given by the sum of the $N_{n}$. This notation should be distinguished from that used to describe a general polytope associated with a (in our case, modified radical) Voronoi tessellation~\cite{Ding2014,Ma2015}. Here $\langle n_{3},n_{4},n_{5},n_{6},\cdots\rangle$ represents a polytope in which $n_{i}$ faces are constructed from $i$ edges, where the labeling $\langle0,0,12,0,\cdots\rangle$ corresponds to the icosahedron. The presently introduced notation labels the icosahedron as $(0,12,0)$. Atoms with $(0,12,Z-12)$ correspond to $Z>12$-coordinated Frank-Kasper polyhedra, whilst $(2,8,0)$, $(3,6,0)$, and $(4,4,0)$ correspond to the Z10, Z9 and Z8 Bernal holes. The Z13 and Z11 Nelson polyhedra correspond to $(1,10,8)$ and $(2,8,1)$. 

Fig.~\ref{FigEvolution}c  plots the $n$-fold bond content evolution during the annealing procedure. Plotted are the 4-, 5- and 6-fold fractional bond populations. Those bonds with non 4-, 5- and 6-fold character are grouped as ``other''. This bond population decreases with relaxation and is at typical values of approximately 0.25\% during the latter stages of relaxation --- it therefore cannot be seen on the vertical linear scale of Fig.~\ref{FigEvolution}c. The population rises by an order of magnitude when using the standard RV tessellation (this is also the case when using the number of edges of the common face to determine $n$). Fig.~\ref{FigEvolution}c demonstrates that the number of defect free 5-fold bonds increases whereas the number of 4-fold and 6-fold defected bond decreases as the glassy structure relaxes. These trends are compatible with the observed increase in $(0,12,0)$ content seen in Fig.~\ref{FigEvolution}b, however the corresponding $n$-fold bond populations change little, indicating that the creation of such low energy and high density icosahedral environments is due to the removal/addition of just a few defect/undefected bonds. Indeed, the $n$-fold bond populations linearly correlate with the icosahedral fractional content, with the gradients equaling -0.26, 0.56 and -0.29 respectively for the 4-, 5- and 6-fold bonds. This indicates that the structural transition associated with the creation of icosahedral content involves (on average) the creation of one 5-fold bond at the expense of a 4-fold and 6-fold defected bond. How this may occur will be discussed in Sec.~\ref{sec:dis}.

\begin{figure*}
	\begin{center}
		\includegraphics[width=0.65\linewidth,trim=1.0cm 0.85cm 1.5cm 0.85cm]{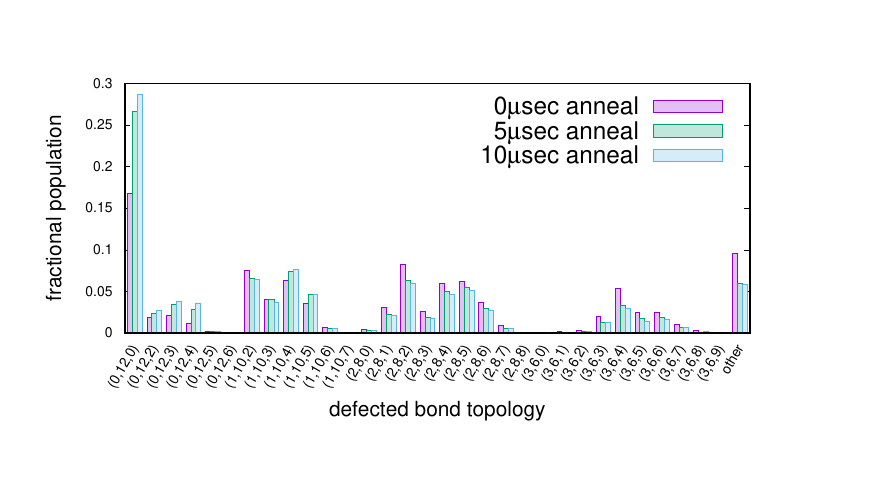}
	\end{center}
	\caption{Histogram of local defected bond structures classed as those satisfying $(N_{4},N_{5},N_{6})=(N_{4},12-2N_{4},N_{6})$ for the 0, 5 and 10 micro-second anneal. Local environments which do not satisfy this are classed as ``other'' and will generally involve non-zero values of $N_{7}$.}
	\label{FigPopulation}
\end{figure*}

Fig.~\ref{FigPopulation} now enumerates the population fractions of those atomic environments completely defined by the $(N_{4},N_{5},N_{6})$ terminology, for the 0 micro-second, 5 micro-second and 10 micro-second anneal. Only non-zero populations are shown, demonstrating that the horizontal axes satisfies $(N_{4},N_{5},N_{6})=(N_{4},12-2N_{4},N_{6})$. This originates from the definition $Z=N_{4}+N_{5}+N_{6}$ and the rule $Z=12-N_{4}+N_{6}$. The Frank-Kasper, Bernal-hole and Nelson polyhedra all satisfy this latter rule, as well as all polyhedra satisfying the SU(2) algebra and originates from the assertion that non-12 coordinated local environments also apply to the SU(2) algebra in which the creation of a 6-fold bond adds an atom to the local environment, and a 4-fold bond removes an atom from the local environment~\cite{Nelson1983b}. There exists an exception to the above rule. The topology $(1,10,1)$ satisfies $(N_{4},N_{5},N_{6})=(N_{4},12-2N_{4},N_{6})$ but is not allowed by the SU(2) algebra of Sec.~\ref{ssec:su2} since (say) $\tilde{\mathbf{U}}^{-}_{1}\tilde{\mathbf{U}}^{+}_{2}\ne\tilde{\mathbf{I}}$, which indicates that it is not possible to terminate a single 4-fold and a single 6-fold defect line at an atomic environment containing no other defect lines. 

Those local environments that could not be classified in terms of a bond topology are grouped into the ``other'' classification. For our least relaxed sample, this is approximately 10\% of the atomic environments, reducing to approximately 5\% for the more relaxed samples. Thus a very large percentage of local atomic environments satisfy the aforementioned bonding topologies.

Fig.~\ref{FigPopulation} reveals that with increasing relaxation, not only do the $(0,12,0)$ icosahedral environments, but also the Frank-Kasper environments increase in number. The Nelson environments involving a single 4-fold bond decrease in number for one and two 6-fold bonds, and increase for three and four 6-fold bonds. All other Nelson environments generally decrease in number. Those environments that do not fit into the SU(2) connectivity picture also decrease in number as the structure becomes more relaxed. Thus generally, a more relaxed structure entails a reduced number of local environments in which many defect bonds connect.

It has to be emphasized that the rule $(N_{4},N_{5},N_{6})=(N_{4},12-2N_{4},N_{6})$ is a realization of the Euler theorem relating the number of faces (bonds) to the number of edges (sum of bond order over bonds) to the number of vertices of the associated modRV polyhedra. This may be revealed through the average number of edges per face via the average bond order defined as:
\begin{equation}
\overline{q}=\frac{\sum_{n} N_{n}\times n}{\sum_{n}N_{n}} \label{eqnedges}
\end{equation}
where $N_{n}$ is the total number of $n$-fold bonds within a given sample. Fig.~\ref{FigEvolution}d plots its time evolution during structural relaxation demonstrating its reduction with time, towards the value associated with the Frank-Kasper crystal phases. Via Euler's theorem this number can be related to the average coordination via
\begin{equation}
\overline{Z}=\frac{12}{6-\overline{q}}, \label{eqncoor}
\end{equation}
indicating that coordination correspondingly reduces with relaxation. Table~\ref{tab1} displays the measured value of $\overline{q}$ and the corresponding prediction for the mean coordination using the above equation. Also shown is the measured mean coordination derived from the number of faces of each atom's Voronoi polyhedron. Data is shown for both the standard RV and modRV, revealing that the modified RV tessellation results in very good agreement with the predictions of Euler. It is noted that when the number of edges of the common face is used to determine the bond order, the Euler relation (Eqn.~\ref{eqncoor}) is followed, with $\overline{q}$. This emphasizes the modification ensures the correct polyehedra connectivity for the common neighbour definition of bond order. Moreover the use of the modified Voronoi tesselation results in $\overline{q}$ values closer to that of the Frank-Kasper phases.
	
\begin{table}[]
	\begin{ruledtabular}
		\begin{tabular}{cccc}
			& measured $\overline{q}$ & $\overline{Z}$ (Eqn.~\ref{eqncoor}) &   measured $\overline{Z}$\\
			\hline
0 $\mu$sec (RV) & 5.200 & 15.005 & 13.872 \\
5 $\mu$sec (RV)  & 5.181 &14.653 &13.720 \\ 
10 $\mu$sec (RV) & 5.175 & 14.553 & 13.762  \\
0 $\mu$sec (modRV) & 5.122 & 13.665 & 13.644 \\
5 $\mu$sec (modRV)  & 5.118 &13.601 &13.470 \\ 
10 $\mu$sec (modRV) & 5.116 & 13.585 & 13.546 
		\end{tabular}
	\end{ruledtabular}
	\caption{Calculated values of the average number of edges per bond ($n$-fold bond value) and the average coordination for samples sample1 and sample2, using the radical Voronoi tessellation and the modified radical Voronoi tessellation introduced in Sec.~\ref{ssec:voro}.}
	\label{tab1}
\end{table}	

\subsection{Application of SU(2) algebra} \label{ssec:su2app}

Whilst the results of the previous section strongly support the notion that the SU(2) bonding rules are largely satisfied in our model binary glass, a quantitative geometric evaluation is still lacking. Fig.~\ref{FigGeneral} demonstrates that not only do their exist rules associated with the $(N_{4},N_{5},N_{6})=(N_{4},12-2N_{4},N_{6})$ classification, there exist relative orinetations between the defect bonds. To quantitatively determine to what extent these very specific $n$-fold bond geometries follow the predictions of the SU(2) algebra (eqn.~\ref{eqnsu2} and Fig.~\ref{FigGeneral}), the local environment of each $(N_{4},12-2N_{4},N_{6})$ is now investigated in detail. 

For each atom, the following procedure is performed:
\begin{enumerate}
	\item The $N_{4}+N_{6}$ defect bond vectors are identified and normalized.
	\item An icosahedron is centred on the central atom and orientated to optimize the alignment of its six five-fold symmetry axes with these bond directions.
	\item Those icosahedron axes closest to the $N_{4}+N_{6}$ bonds are identified with the appropriate $\tilde{\mathbf{U}}^{\pm}_{i}$ operator.
	\item Products of these $N_{4}+N_{6}$ operators which are equal to the identity operator are found to establish that the SU(2) description is obeyed.
\end{enumerate}
The above procedure identifies the SU(2) bond geometry which needs to be minimally distorted to achieve the actual local atomic structure. Application of the procedure to our glass configuration confirms that all $(N_{4},N_{5},N_{6})$ topologies identified in Sec.~\ref{ssec:nfold} do indeed follow the geometry associated with the SU(2) algebra of Sec.~\ref{ssec:su2}. 

\begin{figure*}
\includegraphics[width=0.75\linewidth,trim=1.5cm 3cm 1.5cm 3cm,clip]{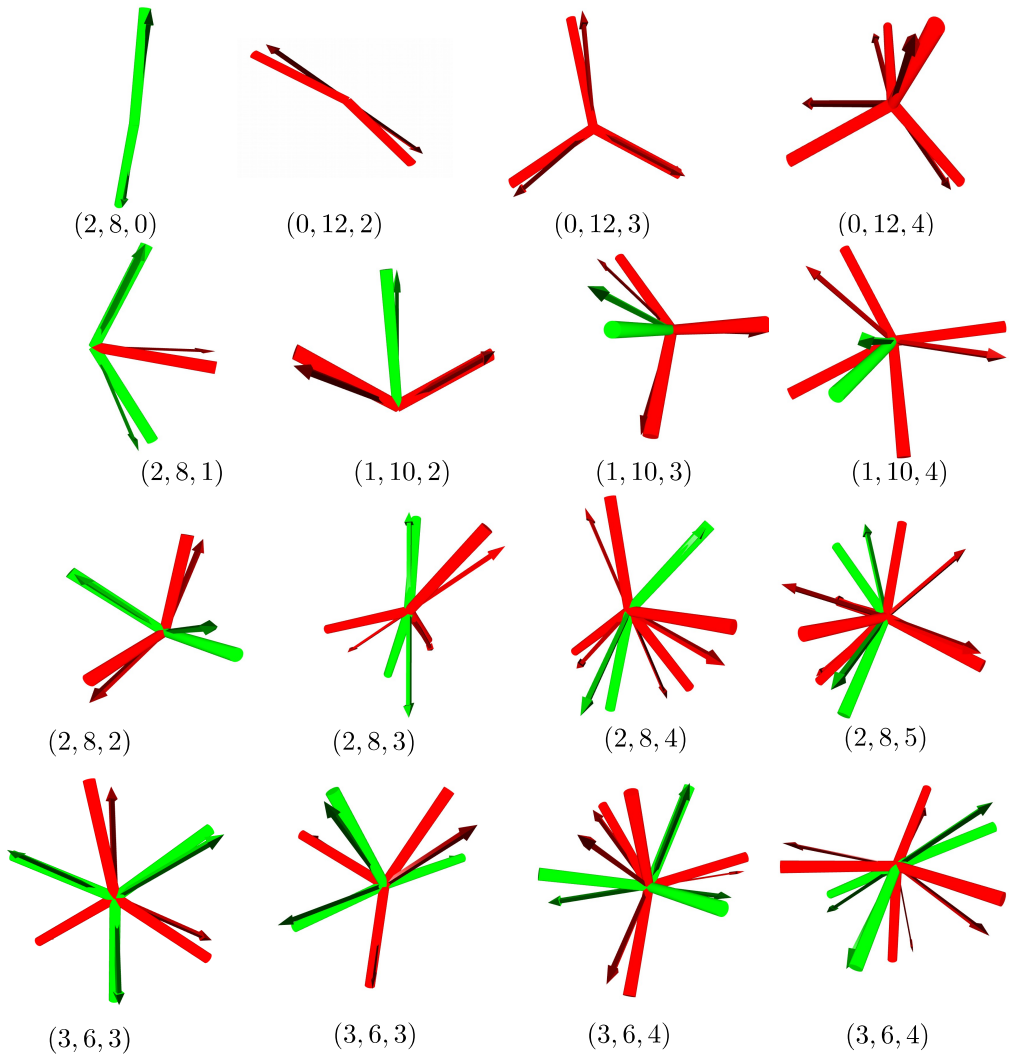}
\caption{Common local defected bond environments derived from the atomistic simulation of a model binary glass. In each panel, the green and red tubes represent the 4-fold and 6-fold atomic bond vectors, whereas the similarly coloured arrows represent the bond vectors associated with to the SU(2) algebra description of Nelson~\cite{Nelson1983a,Nelson1983b}. The upper most row show examples of the Z10 Bernal hole, and the Z14-Z16 Frank-Kasper coordinated polyhedra, whereas the second row show examples of the Z11, Z13-Z15 Nelson coordinated polyhedra. The remaining panels show other commonly occurring defected bond geometries associated with the SU(2) algebra.}
	\label{FigSU2}
\end{figure*}

Fig.~\ref{FigSU2} visualizes some of the more common non-5-fold (defected) bonds for a number of these polyhedra obtained from our most relaxed structure. In the figure, the actual bonds (normalized in length) are visualized as green (4-fold) and red (6-fold) tubes, where as the corresponding 5-fold symmetry axes of the rotated icosahedron are visualized as similarly coloured normalized vectors. In general, correspondence is excellent for the well known bond geometries. It is emphasized that the SU(2) description (the precise angular bond geometry associated with the 5-fold symmetry axes of the icosahedron) is unable do describe local atomic distortions arising from the interaction between atoms of different type --- a certain degree of distortion is thus expected. Both the choice of the orientation (Step 2) and the assignment of operators to the bonds (step 3) minimize this distortion by minimizing the angular difference between the atomic bond vectors and the 5-fold symmetry axes of the icosahedron. The presence of such distortion is best demonstrated for the case of the $(0,12,4)$ Frank-Kasper defected bond geometry (upper right hand panel of Fig.~\ref{FigSU2}) where the four 6-fold bonds arising from the atomic structure exhibit an approximate tetrahedral geometry, a configuration that the 5-fold symmetry axes of the icosahedron cannot describe. To quantify the presence of this distortion, the average distortion defined by the angle between the defect bond obtained from simulation and the relevant 5-fold symmetry axis of the appropriately orientated icosahedron is calculated. Since the minimal angle between 5-fold symmetry axes is approximately 60 degrees, the data shows that the distortion is small where, for environments with 1-3 defect bonds the average distortion is between 5 and 10 degrees. This increases with the number of defect bonds, saturating at appoximately 15 degrees for greater than 5 defect bonds.

The results of the present section therefore establish that atomistic simulations produce a model binary amorphous structure which generally follows the connectivity rules as well as the spatial geometry entailed by the SU(2) algebra formalism of Sec.~\ref{ssec:su2}. In what follows, their relationship to common local structural indicators such as free-volume, energy, stress, and frustration will be investigated.

\subsection{Structural features of SU(2) bond environments and their medium range connectivity}

\subsubsection{Free volume}

Free volume content is a glassy material parameter that can be used to characterize the degree to which an amorphous structure is relaxed, and also how it might respond to an external load. Early theoretical work by Cohen and Turnbull~\cite{Cohen1959} and Spaepen~\cite{Spaepen1977} viewed free volume as well-localized vacancy-type defects, whereas subsequent work views it as a ``diffuse'' de-localized quantity underlying variations in local atomic density~\cite{Argon1979,Egami2011}. Structures with lower free volume (higher densities) tend to be more relaxed. In what follows, a local free volume measure is loosely defined as a deviation of local atomic volume from a reference volume that is presently not defined. In other words, a reduction in local atomic volume would correspond to a decrease in free volume. To assign a volume to a particular atom is somewhat arbitrary, and usually involves either a Voronoi or radical Voronoi tesselation --- or in the present work a modified radical Voronoi tesselation. 

Structurally, regions with high icosahedral content generally contain decreased free volume content. This correlation turns out to be linear suggesting the reduction of free volume and the creation of icosahedral content are one and the same thing~\cite{Derlet2018}. Whilst free volume does not strongly correlate with the spatial occurrence of stress driven athermal structural instabilities (STZ)~\cite{Patinet2016}, regions of reduced free volume tend to exhibit reduced activity involving the thermal activation of localized structural excitations~\cite{Derlet2017,Derlet2018}.

\begin{figure*}
	\begin{center}
\subfloat[]{\includegraphics[width=0.55\linewidth,trim=1cm 0.85cm 1cm 1cm]{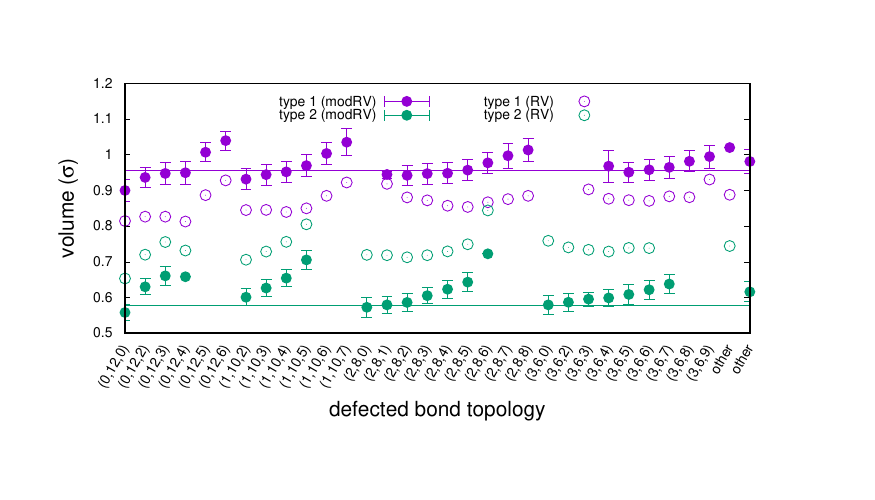}} \\
\subfloat[]{\includegraphics[width=0.55\linewidth,trim=1cm 0.85cm 1cm 1cm]{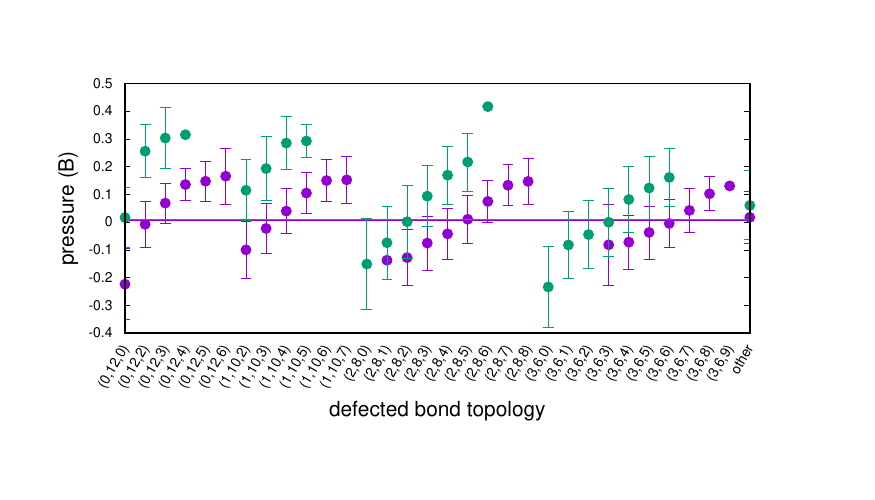}} \\
\subfloat[]{\includegraphics[width=0.565\linewidth,trim=1cm 0.85cm 1cm 1cm]{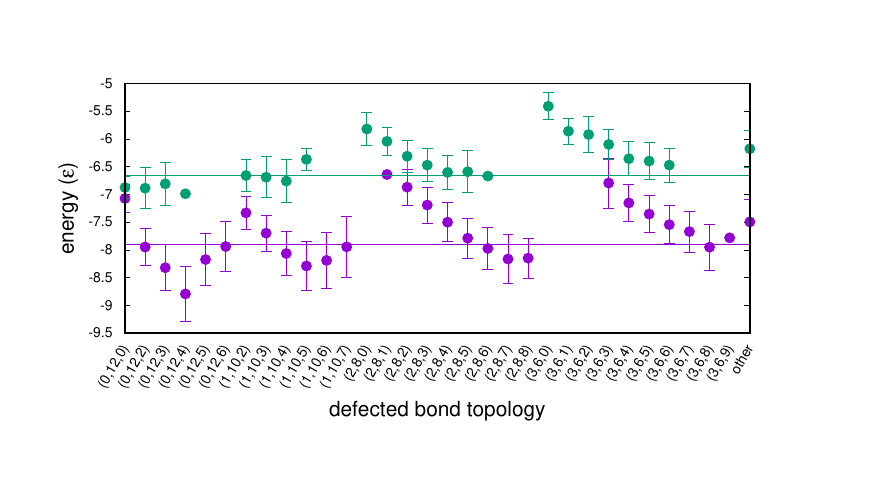}}\\
\subfloat[]{\includegraphics[width=0.55\linewidth,trim=1cm 0.85cm 1cm 0.75cm]{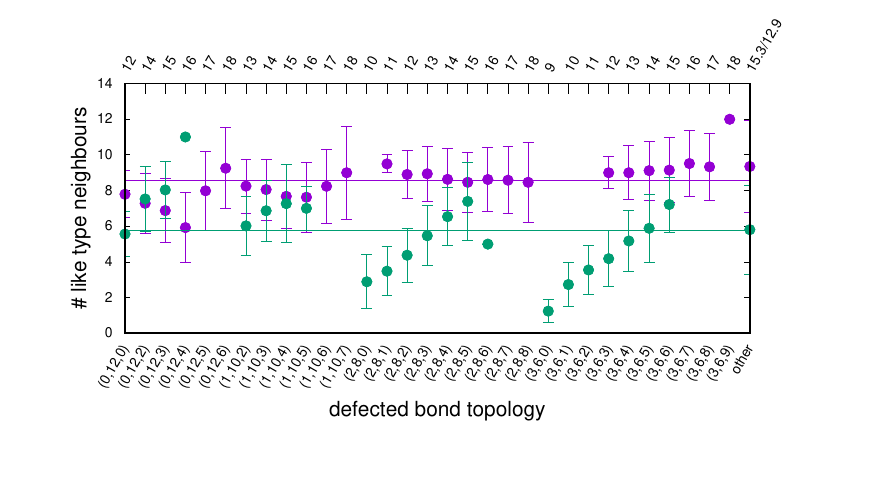}}
	\end{center}
	\caption{Plot of average a) atomic volume, b) local pressure, c) local energy and d) average number of nearest neighbours of the same atomic type, for the local defected bond structures considered in Fig.~\ref{FigPopulation}. Data is shown for the two different atom types using the modified radical Voronoi. For a), the atomic volume data, the standard radical Voronoi tessellation is also used. The lines in each panel indicate the average values for each atom type. In all figures, the error bars are derived from the standard deviation of the scatter. Data is derived from 10 microsecond isotherm configuration. In d), the top x-axis displays the coordination.}
	\label{FigLabelVolZ}
\end{figure*}
 
Fig.~\ref{FigLabelVolZ} plots the average atomic volume for the relevant bond typologies, using both the standard radical Voronoi and modified radical Voronoi tessellations. The classification of the defected bond environments is done via the latter tessellation. Fig.~\ref{FigLabelVolZ}a demonstrates that the resulting average volumes obtained from the modified tessellation lower the atomic volume of the smaller atoms and increase the volume of the larger atom, when compared to the standard tessellation. However, the overall average volume per atom remains the same due to volume conservation, being equal to the total volume of the simulation cell divided by the total number of atoms. Inspection of the figure reveals the well known result that the $(0,12,0)$ topology, the icosahedrally coordinated environment, has the overall lowest volume for each atom type~\cite{Derlet2018}, being well below the global average. The figure also reveals that for each class of bond topologies (distinguished by the number of 4-fold bonds) it is the topology with the lowest number of 6-fold bonds that has the lowest average volume. As the number of 6-fold bonds increases, so does the volume per atom, starting at values below the global average and ending at values above the global average. 

Fig.~\ref{FigLabelVolZ}b now plots the local atomic pressure (divided by a representative bulk modulus $B$ --- see Ref.~\cite{Derlet2012}), showing that regions with reduced volume (high density) are on average under a negative (tensile) pressure, whereas those with enhanced volume (low density regions) are on average under a positive (compressive) pressure. Thus their exists a strong spatial correlation between regions containing low numbers of defect bonds, and regions under tensile pressure which correspond to regions of higher than average density and therefore low free-volume content.

\subsubsection{Local atomic energy}

The work of Ref.~\cite{Derlet2020} found that increased icosahedral content also corresponds to a reduction in the average cohesive energy per atom. From the results of the previous section, one might therefore conclude that spatial regions containing a low bond defect density would also correspond to regions of low cohesive energy. Fig.~\ref{FigLabelVolZ}c shows the average energy per topology class for both atom types. Whilst this conclusion is certainly true of the $(0,12,0)$ environment for the smaller atoms, it is not the case for the larger atoms, where in fact  $(0,12,0)$ environment for the larger atom has one of the highest energies --- a result compatible with the observation that the icosahedral content almost exclusively involves the smaller atom. The conclusion is also valid for the Frank-Kasper topologies for both atom types, where in general the local energy is less than the global average. Indeed, for both atom types, the lowest energy structure is not the bond-defect free topology $(0,12,0)$ but rather the Frank-Kasper Z16 coordinated topology of $(0,12,4)$. This observation most probably drives the nucleation of the Laves phase seen at higher temperatures~\cite{Pedersen2007} and the asymptotic nano-phase structure proposed in Ref.~\cite{Derlet2020}.

For the Nelson topologies, the trend is somewhat more complex. For both atom types, and for a given 4-fold bond number, the energy is highest for the topology containing one 6-fold bond and decreases with the addition of more 6-fold bonds until a minimum is reached beyond which the energy again increases.
For the smaller atom, those local environments containing one 4-fold defect bond, the local cohesive energies are lower than the global average otherwise (for both atom types) it is those Nelson topologies with several 6-fold bonds that are close to or below the global average cohesive energy value. 

The understanding of these trends is difficult because the average atom-type of the nearest neighbour population varies greatly between different realizations of actual atomic environments. Despite the scatter, the average trends are statistically meaningful. Fig.~\ref{FigLabelVolZ}d displays the average number of neighboring atoms of the same type, for both atom types. This figure also plots (on the upper x-axis) the corresponding coordination. The data shows that for the larger atoms, the average population of nearest neighbour large atoms remains approximately constant for all topologies, with the greatest variation occurring within the Frank-Kasper class. For the Nelson topologies, the number of large nearest neighbour atoms remains approximately at eight as the number of 6-fold bonds (coordination) increases. Thus on average, for the larger atoms, adding a 6-fold bond entails an increase in coordination due to the addition of a smaller atom. This trend also appears for the smaller atoms, whose number of small nearest neighbour atoms rises with the addition of a 6-fold bond, with the number of large nearest neighbour atoms staying at approximately eight. For the smaller atom this trend changes at large 6-fold bond numbers, with the addition of a 6-fold bond now corresponding to an additional larger nearest neighbour atom. In Sec.~\ref{sec:dis}, these trends will be discussed in terms of the glass structural models of Miracle and co-workers~\cite{Miracle2003,Miracle2004,Miracle2006,Laws2015}.

Finally, because of the variation of atom type in the nearest neighbour shell, it is not an easy task to conclude via the cohesive energy of an atom, that a low energy glassy structure corresponds to a low density of bond defects. To establish this expected correlation, cohesive energy must be viewed in terms of frustration and the energy that it entails.

\subsubsection{Frustration}

\begin{figure}
\includegraphics[width=0.95\linewidth,trim=1.5cm 0.5cm 1.5cm 0.75cm,clip]{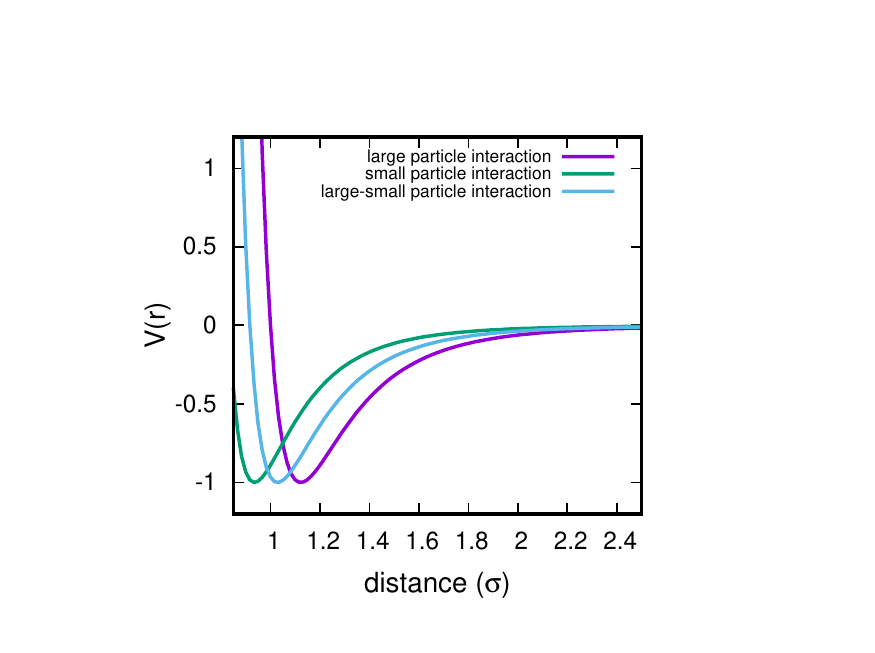}
\caption{Wahnstr\"{o}m parameterization parameterization of the Lennard-Jones interaction between atoms of type 1-1, 2-2 and 1-2, corresponding to the interaction of large, small and large/small atoms. The horizontal axis extends to the employed range of the potential, 2.5$\sigma$.}
\label{FigLJ}
\end{figure}

Structural frustration involves the inability for bonds to mutually achieve a minimum energy (at the equilibrium bond length). Fig.~\ref{FigLJ} plots the radial dependence of the Wahnstr\"{o}m Lennard-Jones (Eqn.~\ref{EqnLJ}) for the three considered interactions. Inspection of the curves reveals the fundamental feature of the Wahnstr\"{o}m parameterization --- that the equilibrium energy between atoms of type $a$ and $b$, which occurs at the radial distance $2^{1/6}\sigma_{ab}$, is equal to $-\varepsilon$ and thus independent of atom type --- the energy of the unfrustrated bond is independent of atom size. 

Deviation from the distance $2^{1/6}\sigma_{ab}$ is a direct measure of the geometrical frustration. Defining the frustration as the fractional deviation away from the equilibrium distance:
\begin{equation}
f_{ab}(r)=\frac{r-2^{1/6}\sigma_{ab}}{2^{1/6}\sigma_{ab}}, \label{EqnGF}
\end{equation} 
the Wahlstrom Lennard Jones interaction becomes
\begin{equation}
V(f(r))=\varepsilon\frac{1-2\left(1+f_{ab}(r)\right)^{6}}{\left(1+f_{ab}(r)\right)^{12}} \label{EqnVLJGF}
\end{equation} 
where $r$ is the actual bond length. Thus the energy increase due to the defined geometrical frustration measure is also independent of atom type.

\begin{figure*}
	\begin{center}
\subfloat[]{\includegraphics[width=0.55\linewidth,trim=1cm 0.85cm 1cm 1cm]{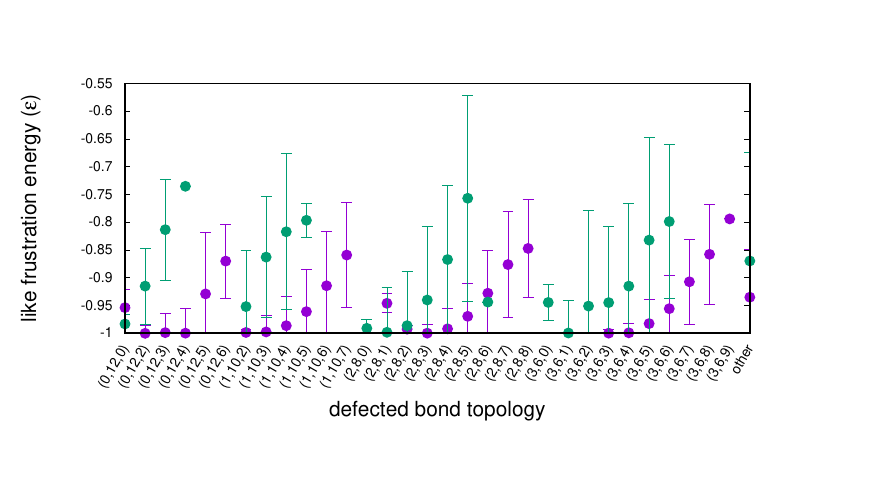}} \\
\subfloat[]{\includegraphics[width=0.55\linewidth,trim=1cm 0.85cm 1cm 1cm]{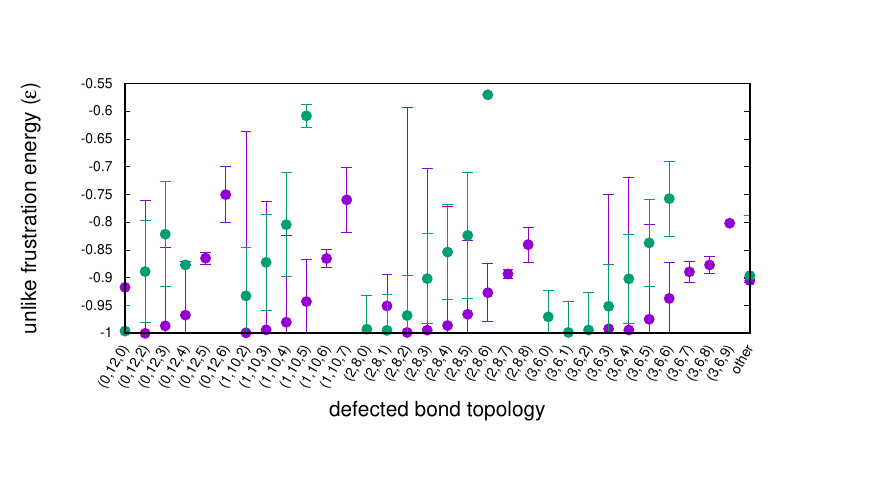}} \\
\subfloat[]{\includegraphics[width=0.55\linewidth,trim=1cm 0.85cm 1cm 1cm]{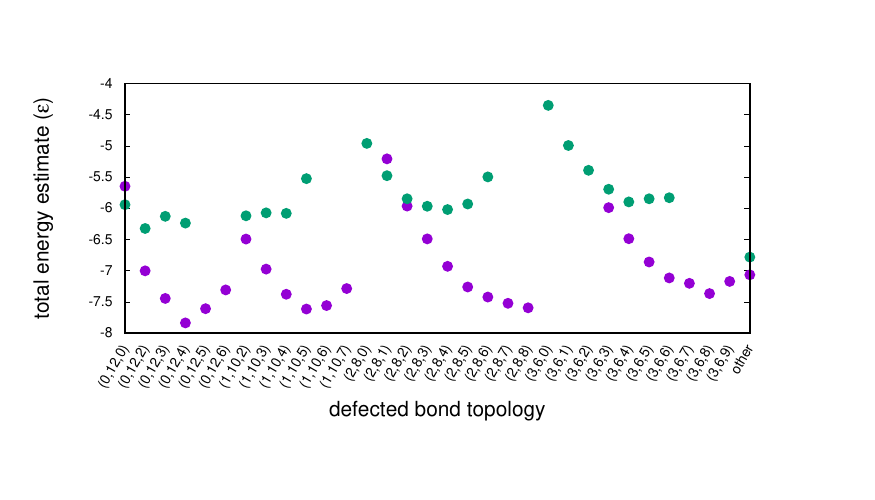}}
	\end{center}
	\caption{Plot of average a) like and b) unlike energy of frustration (Eqn.~\ref{EqnVLJGF}), using the average geometrical frustration (Eqn.~\ref{EqnGF}) of each topology environment. c) Shows the corresponding estimates of the local cohesive energy derived from the data of a) and b), and Fig.~\ref{FigLabelVolZ}d. Data is derived from 10 microsecond isotherm configuration.}
	\label{FigLabelBondLength}
\end{figure*}

Using Eqn.~\ref{EqnGF}, the average geometrical frustration of like and un-like atoms may be calculated for both atom types, as a function of defect bond topology. These average geometrical frustration values may then be inserted into Eqn.~\ref{EqnVLJGF} to calculate the corresponding energy of frustration. Figs.~\ref{FigLabelBondLength}a-b do this and directly show that for both atom types, those local topologies with a minimum number of bond defects generally have the lowest energy of frustration. Where this is sometimes not the case, is in the regime when the number of 4-fold bonds is greater than or equal to the number of 6-fold bonds (for example the $(2,8,1)$ for the larger atom and the $(3,6,0)$ for the smaller atoms). Because of their low coordination, such environments are however quite rare ($\lesssim0.2\%$ of the identified topologies). 

Figs.~\ref{FigLabelBondLength}a-b establish numerically, that a glassy structure with a reduced bond-defect content does indeed correspond to a minimally frustrated low energy structure. To connect this data to the local cohesive energy averages shown in Fig.~\ref{FigLabelVolZ}c, an estimate of the total energy using the frustration energy entailed in Figs.~\ref{FigLabelBondLength}a-b and the average nearest neighbour occupancies of Fig.~\ref{FigLabelVolZ}d can be made. This is shown in Fig.~\ref{FigLabelBondLength}c, which is qualitatively similar to that of the actual local cohesive energy averages calculated using the entire range of the Lennard-Jones interaction (Fig.~\ref{FigLabelVolZ}c). Indeed, detailed comparison indicates that the effect of more distant interactions is mainly characterized by a global energy offset, suggesting that these contribute to the energetics in mean-field-like way.

\subsubsection{Spatial structure}

The pair distribution function (PDF) gives some insight into the short and medium range atomic structure of a glass, and is indirectly accessible via x-ray diffraction. It is generally characterized by a strong peak at distances comparable to the nearest neighbour separation between atom, showing that like that of the liquid, the glass has strong short range order. At longer distances comparable to the second nearest neighbour distances, a very broad peak is seen which (for well relaxed glasses) has some structure indicating medium range order (MRO). Such MRO is difficult to see in atomistic simulation due to the short time-scales simulated, however the present micro-second scale simulation do result in a sufficiently relaxed amorphous structure to reveal this MRO signature~\cite{Derlet2020}. 

\begin{figure}
	\begin{center}
\subfloat[]{\includegraphics[width=0.95\linewidth,trim=1cm 0.5cm 1cm 1cm]{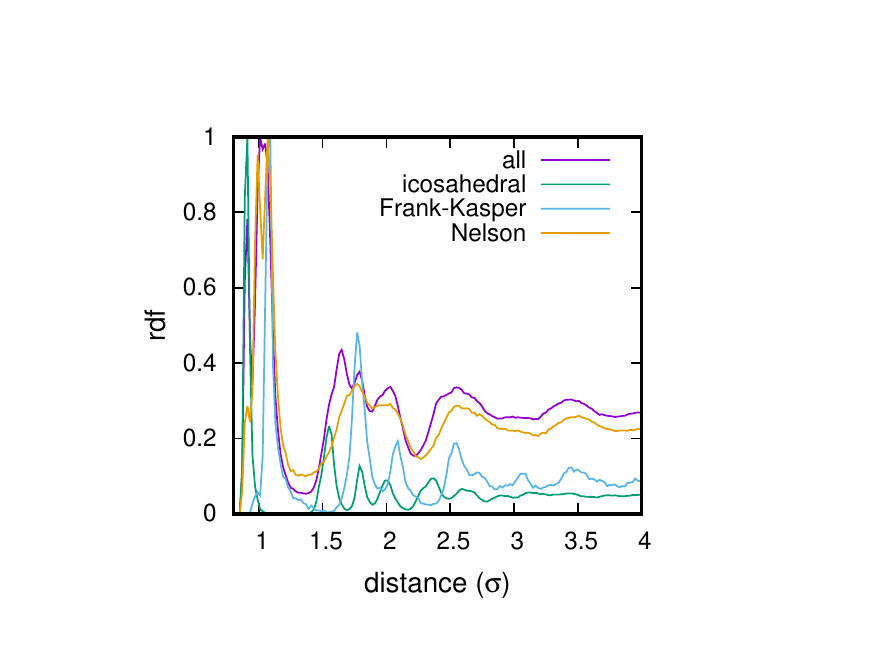}} \\
\subfloat[]{\includegraphics[width=0.95\linewidth,trim=1cm 0.5cm 1cm 1cm]{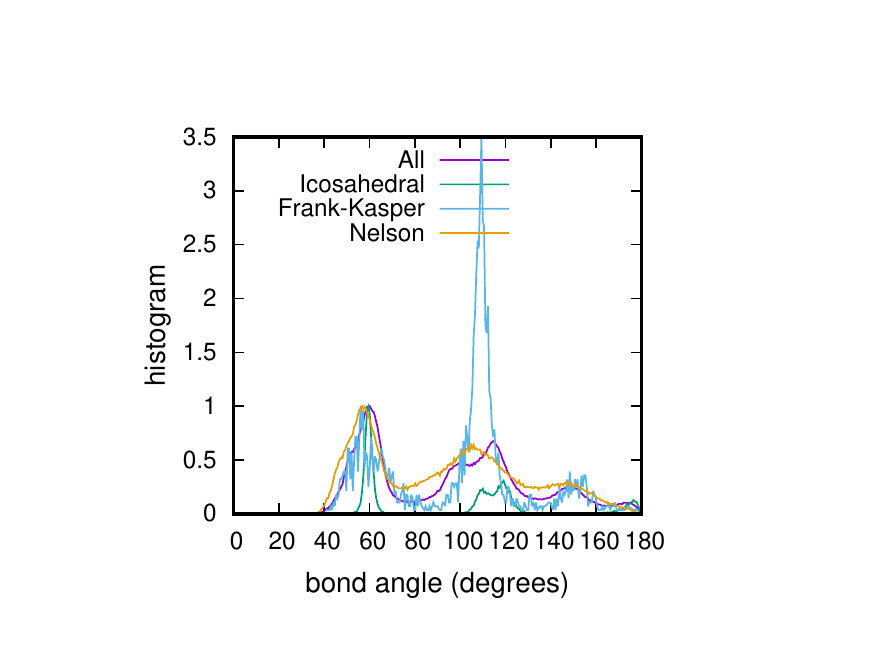}}	\end{center}
	\caption{a) Pair and b) bond angle distribution of sample 2 after 10 $\mu$secs of isothermal annealing, showing the total distribution arising from all atoms, and corresponding partial distributions arising from atoms in icosahedral, Frank-Kasper and Nelson environments. In both figures, each distribution has its first peak height normalised.}
	\label{FigPDF}
\end{figure}

Fig.~\ref{FigPDF}a plots the PDF for the 10 $\mu$sec sample, indicating the usual strong SRO peak at approximately a bond length ($\sim\sigma$) and a more distant MRO structure in the second peak. Inspection of the partial PDFs due to icosahedral ($(0,12,0)$), Frank-Kasper ($(0,12,2)$, $(0,12,3)$, etc.) and Nelson ($(1,10,2)$, $(1,10,3)$, ...) environments are also shown, and indicate the origin of the MRO seen in the total PDF. Indeed, the MRO can be largely understood in terms of the icosahedral and Frank-Kasper PDFs, which both have a very clear peak structure that arises directly from the SU(2) angular relations of the 6-fold bonds and which is similar to that of the C15 Laves polyhedra structure. For the icosahedral environments it is mainly the smaller atom type which is involved ($\sim95\%$), whereas for the Frank-Kasper environment it is the larger atom ($\sim95\%$). On the other hand the Nelson environments follow more closely the total PDF, with approximately equal concentrations of small and large atoms, suggesting their distribution is more random. As the glassy structure becomes more relaxed, an increased fine structure becomes apparent that extends well beyond the second nearest neighbour regime and is largely due to more distant correlations between the icosahedral and Frank-Kasper environments.

These trends are also reflected in the icosahedral, Frank-Kasper and Nelson angular distribution functions shown in Fig.~\ref{FigPDF}b, for which the icosahedral and Frank-Kasper curves show distinct peaks. The peaked regions of the correlation functions are at angular values which are again close to the geometries predicted by the SU(2) relations and that seen in the C15 Laves structure, and again the Nelson environments follow the angular distribution function arising from the average of all atomic environments. The degree of structural relaxation is best reflected in the peak at approximately 120 degree, which correspondingly increases. This angle reflects the increasing presence of  $(0,12,3)$ and $(0,12,4)$ environments, as seen in the populations of Fig.~\ref{FigPopulation}.

\begin{figure*}
	\begin{center}
\subfloat[]{\includegraphics[width=0.45\linewidth,trim=1.5cm 0.5cm 1.5cm 1cm]{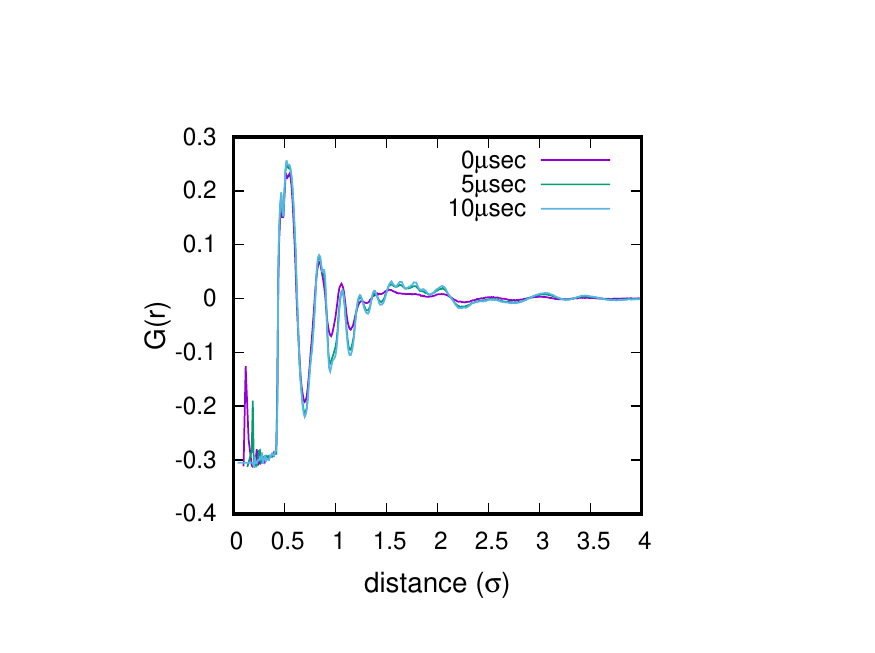}} 
\subfloat[]{\includegraphics[width=0.45\linewidth,trim=1.5cm 0.5cm 1.5cm 1cm]{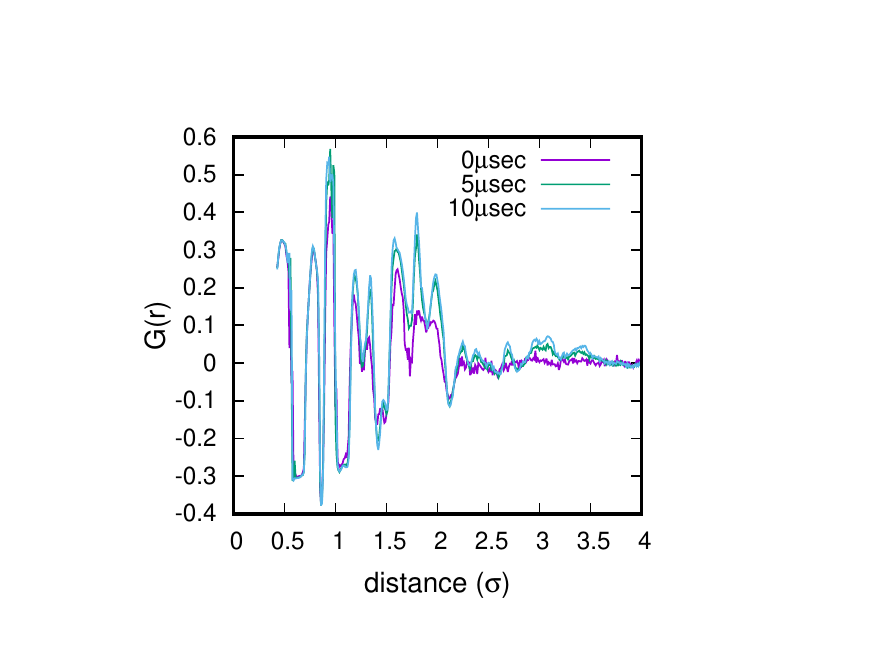}}\\
\subfloat[]{\includegraphics[width=0.45\linewidth,trim=1.5cm 0.5cm 1.5cm 1cm]{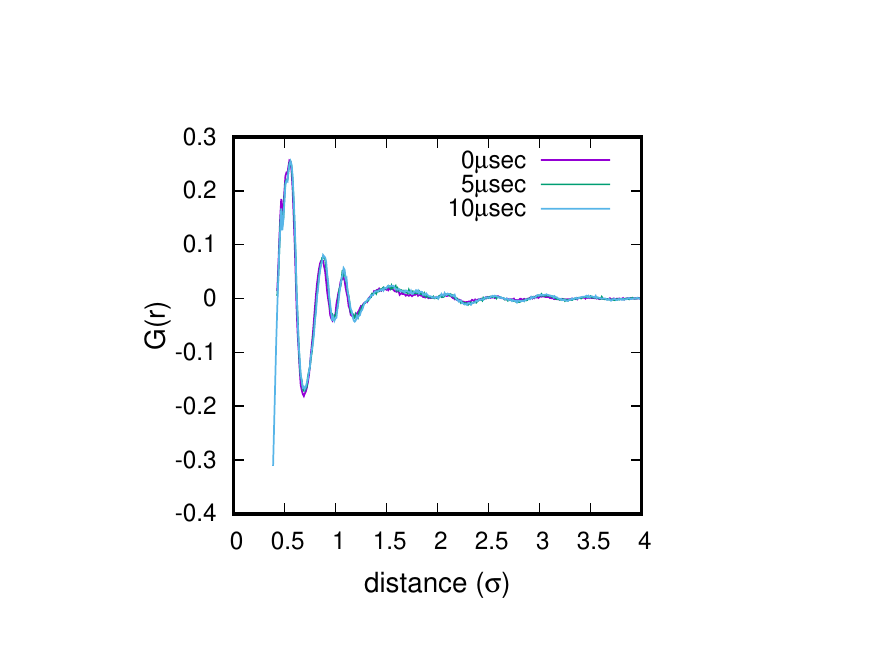}} 
\subfloat[]{\includegraphics[width=0.45\linewidth,trim=1.5cm 0.5cm 1.5cm 1cm]{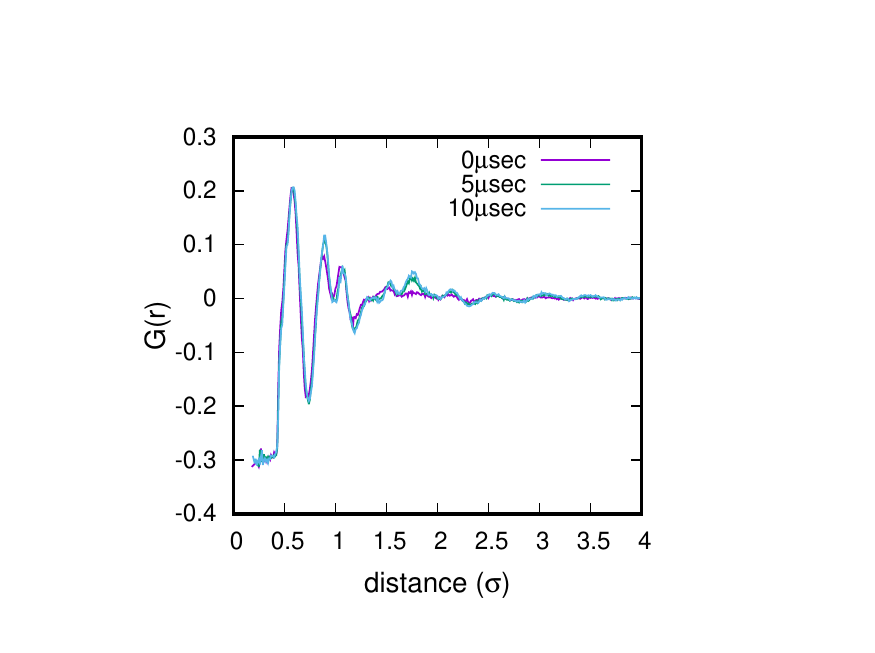}}
	\end{center}
	\caption{Plot of the orientational correlation function (Eqn.~\ref{EqnCF})  derived from the four bond groupings: a) all bonds, b) 5-fold bonds involved in $(0,12,0)$ atomic environments, c) the other remaining 5-fold bonds, and d) defected 4-fold and 6-fold bonds for the 0, 5 and 10 microsecond relaxation isotherms.}
	\label{FigQQ}
\end{figure*}

As in past work, this demonstrates that the MRO of this model glass is characterized by fragments of bond networks whose geometry is similar to the polyhedral backbone of the C15 Laves structure. It is noted that the structure of the C15 Laves crystal is characterized by an extended array of 6-fold defect bonds mediated by a connected network of Frank-Kasper $(0,12,4)$ local geometries. This is interpenetrated by a network of defect free bonds. For the case of our glass, fragments of this type will require a population of $(0,12,2)$ and $(0,12,3)$ local geometries to accommodate the disorder. Indeed, the present work demonstrates this to originate from two distinct contributions, that of 5-fold bond fragments underlying the icosahedral defect free structures and that of 6-fold bonds underlying Frank-Kasper defect lines characterized by the $(0,12,2)$, $(0,12,3)$, $(0,12,4)$, ... local topological environments. Bond connectivity between these two networks is via 4-fold bond defect lines whose termination is described by the Nelson polyhedra and the rules of the SU(2) algebra. This later environment appears more random due to the large configurational possibilities of the 5-fold and 6-fold fragments of the C15 structure. These observations are entirely compatible with population histograms shown in Fig.~\ref{FigPopulation} and demonstrate the more general picture given by the SU(2) formalism.

To quantify the extent of spatial correlation of both the defect free and defected environments a spatial correlation function that is sensitive to deviations away from a global 5-fold icosahedral symmetry is required. Following Refs.~\cite{Steinhardt1981,Steinhardt1983,Nelson1983b} this may be achieved through defining for each bond:
\begin{equation}
Q_{6m}(\mathbf{r})=Y_{6m}(\theta(\mathbf{r}),\phi(\mathbf{r})),
\end{equation}
where $\mathbf{r}$ is the location of the mid-way point of the bond, and $\theta(\mathbf{r})$ and $\phi(\mathbf{r})$ are its polar and azimuthal orientation with respect to an (arbitrary) global coordinate system. $Y^{lm}(\theta,\phi)$ is the $l=6,m=-l,\cdots, l$ spherical harmonic (see table II of Ref.~\cite{Nelson1983b}). The corresponding rotationally invariant icosahedral correlation function is
\begin{equation}
G(\Delta\mathbf{r})=\frac{4\pi}{13}\sum_{m=-6}^{6}\left\langle Q_{6m}(\mathbf{r})Q_{6m}(\mathbf{r}+\Delta\mathbf{r})\right\rangle, \label{EqnCF}
\end{equation}
where $\langle\cdots\rangle$ represents an average over the glassy micro-structure. This correlation function measures the average degree of orientational icosahedral alignment between two bonds separated by a distance $\Delta\mathbf{r}$. Due to a global structural isotropy, $G(|\Delta\mathbf{r}|)$ is only considered. 

Fig.~\ref{FigQQ} plots the obtained correlation functions for the 0, 5 and 10 microsecond configurations using four different bond groupings: a) all bonds, b) 5-fold bonds involved in $(0,12,0)$ atomic environments, c) all other remaining 5-fold bonds, and d) all defected 4-fold and 6-fold bonds. For the latter two cases, bonds that belong to atomic environments not identified by the SU(2) formalism are omitted. Doing so, results in somewhat sharper features for the correlation functions. Fig.~\ref{FigQQ} demonstrates all correlation functions feature three distinct peaks at approximately 0.5$\sigma$, 0.8$\sigma$ and $\sigma$. These correspond to bond separations within the environment of a single atom assuming an icosahedral geometry which will be the case of the $(0,12,0)$ environment and also that of the 4-fold and 6-fold bonds identified through the SU(2) analysis. The correlation of 5-fold bonds between $(0,12,0)$ environments (Fig.~\ref{FigQQ}b) exhibits a strong correlation up to distance of two atomic bond lengths ($\approx 2\sigma$). As the structure becomes more relaxed, a fine structure emerges at separations between $\sigma$ and 3$\sigma$ which is mainly due to second nearest neighbour correlations between $(0,12,0)$ environments (Fig.~\ref{FigQQ}b) and also the 4-fold and 6-fold defect bonds (Fig.~\ref{FigQQ}d). Such correlations, albeit reduced, also exist at further distances up to approximately 4-5$\sigma$. It is noted that when sample size statistics as a function of $|\Delta\mathbf{r}|$ have been corrected for, no long-range asymptotic form such as 
$\sim\exp(-r/\xi)$ was observed. Thus the exponential envelope observed in the work of Steinhardt {\em et al}~\cite{Steinhardt1981,Steinhardt1983} where $\xi$ was observed to be 3-5 particle diameters (for a mono-atomic LJ model) was not observed.

\begin{figure}
	\begin{center}
\includegraphics[width=0.9\linewidth,trim=4.75cm 1cm 4cm 1cm,clip]{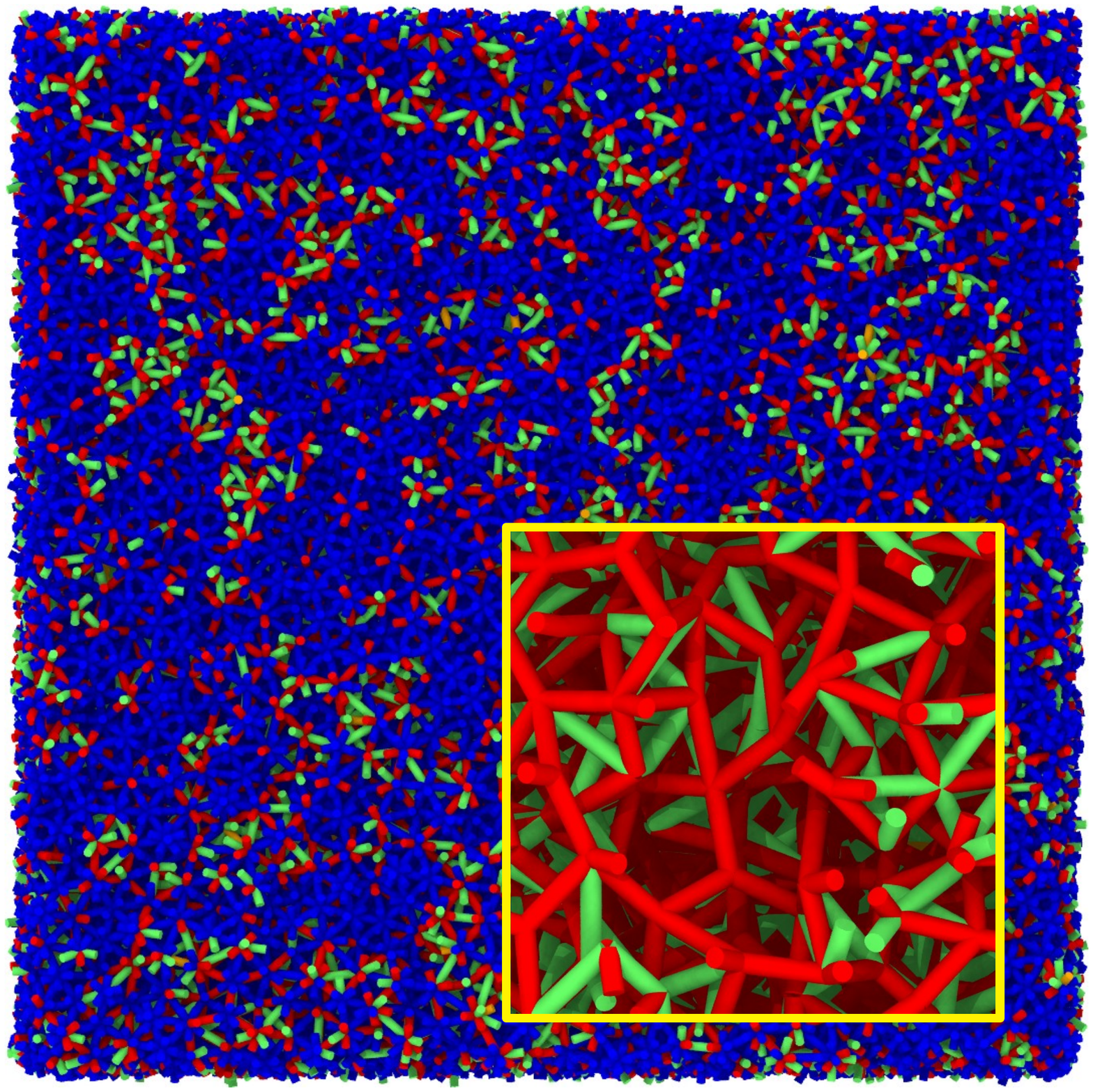}
		\caption{a) Visualization of all bonds of sample2, coloured according to their $n$-fold value (4-green,5-blue,6-red, with orange indicating other $n$ values). Inset displays a region showing only defected bonds --- the 4-fold and 6-fold disclination line network.}
		\label{FigVis}
	\end{center}
\end{figure}

The extended spatial structure of these three region types is visualized in Fig.~\ref{FigVis} for the 10 microsecond sample in which the 4-, 5-, and 6-fold bonds are coloured according to the usual green, blue and red convention of Figs.~\ref{FigGeneral} and \ref{FigSU2}. Fig.~\ref{FigVis} reveals a connectivity involving system spanning 4-fold, 5-fold and 6-fold bond clusters forming an extended network of disclination lines. For the case of the 5-fold and 6-fold cluster, almost all bonds belong to their respective super-cluster, whereas for the 4-fold bonds, there exists one system spanning cluster and several hundred smaller clusters of up to 100 4-fold bonds. This results in all atoms being connected by the 5-fold bond super-cluster, and approximately 75\% of the atoms being connected by 6-fold bonds (all of the larger atoms and 50\% of the smaller atoms). Close inspection of the disclination network reveals that the 4-fold and 6-fold bonds exhibits connected fragments of the Laves C15 backbone polyhedra --- the Frank-Kasper topologies of $(0,12,2)$, $(0,12,3)$, $(0,12,4)$. For the 6-fold bonds, this is shown in the inset which visualizes a smaller portion of the system without the defect free 5-fold bonds. The inset reveals a mixture of Frank-Kasper and Nelson topologies and presents a realization of the defect structure of the amorphous solid. 

Fig.~\ref{FigVis} should be compared to the schematic originally proposed by Nelson~\cite{Nelson1983a,Nelson1983b,NelsonBook}, which shows a defect network with a considerably lower line-defect density. In this schematic, so-called isolated bubble structures are shown consisting of two nearby $(1,12,2)$ topologies. Since both of these have a coordination of Z=13, and therefore share an extra atom, such a ``bubble'' defect was referred to as an intersitial defect. A similar structure involving nearby $(2,8,1)$ topologies was referred to as a vacancy defect. Such isolated structures, which should involve at least 4 atoms, where not observed in the atomic configurations of the current work. The $(1,12,2)$ tended more to be at the boundary of between regions of defects and defect-free regions.

\section{Discussion} \label{sec:dis}

In 1952, Frank asked the question~\cite{Frank1952} ``In how many different ways can one put twelve billiard balls in simultaneous contact with one, counting as different the arrangements which cannot be transformed into each other with out breaking contact with the centre ball?''. The answer is three, the two close-packed configurations of the face-centered cubic and hexagonal structures, and the icosahedron structure. For soft inter-atomic potential systems, this hard-sphere constraint entails a non-negligible barrier energy separates the three configurations, if the allowed transformation trajectories are constrained to a certain distance from the central atom. Relaxing this constraint by allowing the particles to move away from the central atom, can reduce this barrier energy. If these thirteen atoms are not alone, such as in a densely packed structure of a glass, this structural trajectory becomes limited, allowing one to conclude that generally the three configurations of Frank are separated by non-negligible energy barriers. A similar rational exists for the SU(2) defect structures of Nelson --- energy barriers must be crossed for there to be a re-organisation of the nearest-neighbour shell defect structure. Moreover, since this latter aspect involves topological defects within the structural degrees of freedom of the nearest neighbour shell, the corresponding barrier energies are expected to be large. It is from this perspective that one must view the present formalism as a low-energy description of nearest neighbour structure with the obvious caveat that it is an approximate theory.

The original discussion by Frank, and indeed that of Nelson, explicitly considered similar sized atoms and the present work suggests it may be extended to atoms of different size, specifically for atomic radii ratios of about $R_{\mathrm{small}}/R_{\mathrm{large}}=5/6\approx0.83$. How might the present results vary across wider ranges of atomic size differences? Insight into this question can be found in the work of Miracle and co-workers~\cite{Miracle2003} who considered the efficient packing of the first nearest neighbour shell of differently sized atoms. In particular they considered the most efficient packing of solvent atoms around a central solute atom as a function  $R_{\mathrm{solute}}/R_{\mathrm{solvent}}$ and found that depending on the value of this ratio, a particular choice of ``surface'' coordination was needed to better understand the experimental trends. They considered the regimes of $n=3$, 4, 5, 6 and higher surface coordinations, the latter three of which would correspond to the strain free efficient packing of only 4-, 5- and 6-fold bonds. For the $n=3$, 4 and 5 surface coordinations these were viewed as tetrahedral, octahedral and icosahedral arrangements of solvent atoms around a solute atom. The work found that specific values of $R_{\mathrm{solute}}/R_{\mathrm{solvent}}$ would entail these quite different geometrical arrangements corresponding to respectively 0.225, 0.414 and 0.904. For radii ratios between these values, a less efficient but geometrically similar packing was envisaged.

\begin{figure}
	\begin{center}
\includegraphics[width=0.95\linewidth,trim=1cm 0.85cm 1cm 0.75cm]{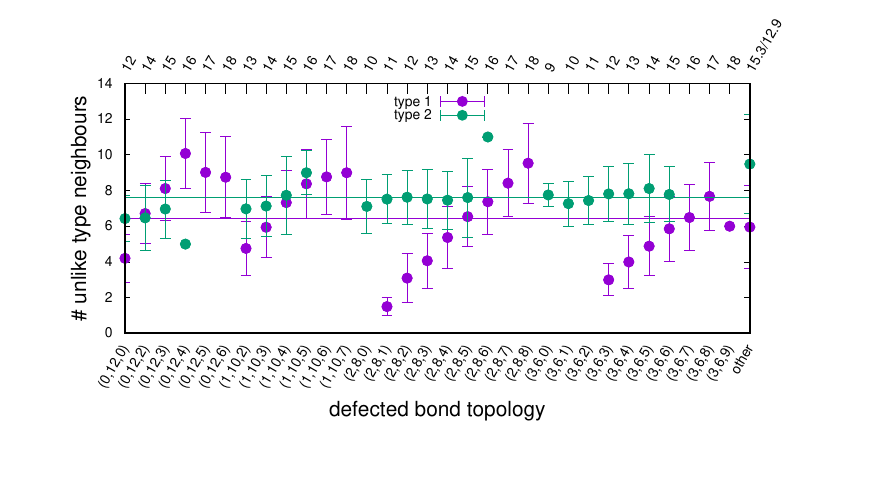}
	\end{center}
	\caption{Plot of average average number of nearest neighbours of the different atomic type, for the local defected bond structures considered in Fig.~\ref{FigPopulation}. Data is a replot of Fig.~\ref{FigLabelVolZ}d}
	\label{FigLabelCID}
\end{figure}

To relate the present work to the findings of Ref.~\cite{Miracle2003}, Fig.~\ref{FigLabelCID} replots the data of Fig.~\ref{FigLabelVolZ}d in terms of the number of un-like atoms in the nearest neighbour shell. For the Nelson polyhedra, both the smaller and larger atoms have an approximately constant average number of larger atom nearest neighbours, with the average number of smaller nearest neighbour atoms increasing with the addition of a 6-fold defect bond. When considering this trend from the smaller atom environment (that is, viewing the smaller atom as the solvent atom), the average number of large atoms is approximately eight with the upper error bar range reaching between nine and ten nearest neighbours. These numbers are close to, but below the maximum packing efficiency for a surface coordination regime of 4, predicted by Miracle and co-workers~\cite{Miracle2003} which is approximately 10.5 for our atomic radii ratio of $R_{\mathrm{small}}/R_{\mathrm{large}}=5/6\approx0.83$. Such a ratio does does not allow for a packing of the icosahedral structure with only largest atoms in the nearest neighbour shell ---  a result compatible with the observation that the $(0,12,0)$ environments on average have equal populations of small and large nearest neighbor atoms. Within each class of Nelson topologies, $(0,10,.)$, $(2,8,.)$ and $(3,6,.)$, the mean bond order starts respectively with values of 5.1, 4.8 and 4.6 and then for all three classes rises to a value of approximately 5.3 as the 6-fold bond defect number increases to its maximum observed value. Whilst the scatter is large, the larger atom is least likely to have a 6-fold coordination, with a comparable chance of being either having a 4- or 5-fold coordination. This trend is compatible with the surface coordination of 4 predicted by Ref.~\cite{Miracle2003}. Generally, a 6-fold bond is more likely to involve a smaller atom for both central atom types, a somewhat intuitive result since it should be easier to pack atoms around a bond containing smaller atoms. 

The above results demonstrate a compatibility with the glass structural model of Ref.~\cite{Miracle2003} and suggest that changing the values of $R_{\mathrm{small}}/R_{\mathrm{large}}$ will change the populations of 4-, 5- and 6-fold bonds, and thus the icosahedral population within the binary glass structure. Moreover in the extreme limit of this ratio becoming very small or very large, the present description (in terms of the five-fold symmetry of the icosahedron) might break down due to the presence of 3-fold and 7-fold bonds. Such regimes of radii ratios are however not common experimentally. The work of Ref.~\cite{Miracle2003} has been extended to overlapping nearest neighbour shells, producing the so-called efficient cluster packing model of glass structure and a theory of MRO~\cite{Miracle2004,Miracle2006,Laws2015}. It is an interesting prospect that the present SU(2) theory of bond-defects might facilitate a more precise realization of such packing models.

When one considers several neigbouring nearest neighbour environments, in terms of start and end configurations, the admitted structural transformation among them (which may or may not involve a change in coordination) should ultimately follow the local SU(2) connectivity rules. However it is not given that the corresponding energy barriers between such configurations be large. Indeed they may be arbitrarily small. For example small displacements resulting in an atom (or atoms) changing which atoms constitute its neighbours, can result in a change in the local topology of the nearby atoms without necessarily passing through a large energy barrier. To investigate this aspect atomistic configurations at different times are investigated.

\begin{figure*}
	\begin{center}
		\includegraphics[width=0.95\linewidth,trim=1cm 2cm 1cm 2cm]{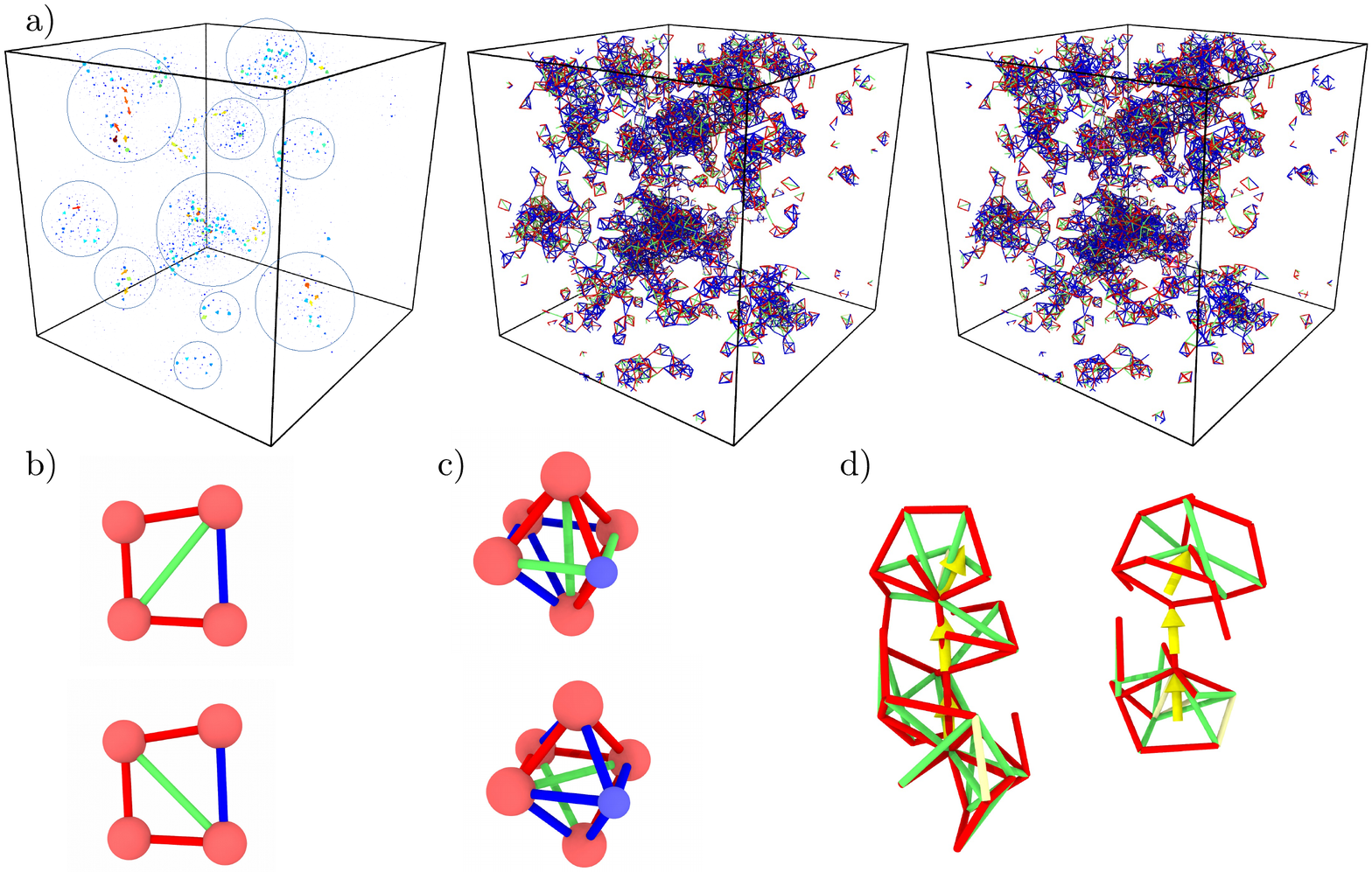}	\end{center}
	\caption{a) Visualization of atomic displacement field occurring between two configuration separated by 10 nano-seconds. Displacement vectors are colour red for displacement magnitudes comparable to the typical bond length and blue for displacement magnitudes associated with non-affine strain. A number of localized structural excitations are circled, in which neighbour atoms successively displace.}
	\label{FigChange}
\end{figure*}

Fig.~\ref{FigChange}a visualizes an example of thermally activated structural change occurring between two configuration separated in time by 10 pico-seconds at the 14 microsecond timescale of the isotherm anneal. The left-most panel plots the corresponding atomic displacement vectors of, and colours them according to their magnitude. Red indicates the largest displacements, which is are comparable to a bond length, and represent localized structural excitations (LSEs)~\cite{Swayamjyoti2014,Swayamjyoti2016,Derlet2017}. On the other hand, the shorter blue displacement vectors indicate the smallest scale of displacement and generally correspond to the surrounding accommodating strain which accompanies the LSE. LSE structure generally involves bond-length scale displacements involving a few atoms, in which neighbouring atoms replace each others position forming a string/loop like displacement structure. The center and right most panels of Fig.~\ref{FigChange}a now show only the change in the $n$-fold bond structure. Here, only atoms whose local topology has changed are displayed. The bond structure (at the start and end) connecting these atoms is also displayed. A broad inspection of the spatial structure shows strong overlap with the identified LSEs of the left panel --- an obvious result reflecting the strong change in the local bonding environment due to the displacements associated with the LSEs.

In regions away from the LSEs, where there only exist small displacement fields due to the accommodating strain fields of the LSEs, changes in the local bonding topology are largely absent. In these parts of the structure, there do however exist localized and isolated changes in the bonding topology. Figs.~\ref{FigChange}b and c, display a zoom-in of some common examples of such changes. Here the upper and lower panel of b) and c) show the configurations at the start and end of the 10 pico-second interval. In b), which displays a planar structure, a 4-fold bond has switched between two different pairs of atoms with no other change in bond structure between the displayed atoms. On the other hand, c) which displays a sextet of atoms associated with a 4-fold bond indicates a switching of this 4-fold bond between two different atom pairs. In this case, the switching results in a different bonding structure between the surrounding common-neighbour atoms.

Given that the above examples occur in regions experiencing minor non-affine displacements, due to strain, one might simply attribute such fluctuations as a result of the choice of the criterion of what constitutes a nearest neighbour bond. This is a generic problem of metallic systems due to the de-localized nature of the atomic bonding, relegating the choice of atom size and bonding connectivity to a non-unique choice of local geometry such as in the Voronoi and radical-Voronoi tessellations. Alternatively, this ambiguity does reflect a type of degeneracy in the topology which also corresponds to an obvious approximate degeneracy in energy when applying Eqns.~\ref{EqnGF} and \ref{EqnVLJGF} to each identified bond (as was done in Figs.~\ref{FigLabelBondLength}).

In regions where LSEs occur there exist strong changes in the nearest neighbour structure and therefore in the $n$-fold bonding structure. Fig.~\ref{FigChange}d gives one example of this where the LSE is characterized by three atoms linearly displacing (indicated by the yellow arrows). Surrounding this LSE is a displacement field that reduces in magnitude as the distance from the central LSE increases (not shown). The left and right panels in d) show the non-5-fold structure associated with the nearest neighbours of the three-atom LSE, at the start and end of the 10 pico-second interval. The local structure of the lower atom contains non 4- and 6-fold bonds, and therefore is in an environment not straight-forwardly characterized by the SU(2) topology of Sec.~\ref{ssec:su2}. On the other hand, the central atom transits from a Nelson $(2,8,5)$ topology to a Frank-Kasper $(0,12,3)$ topology, whereas the upper atom whose initial 4- and 6-fold environment could not be uniquely identified has transited to the Nelson $(2,8,2)$ topology. Thus the LSE has resulted in a reduction of the local defect structure and the creation of a Frank-Kasper 6-fold structure.

More generally, the above analysis suggests a topological reason for the observed string-like geometry of the LSEs (which are also seen within the dynamical heterogeneities of the under-cooled liquid~\cite{Donati1998,Schroeder2000,Gebremichael2004,Vogel2004,Kawasaki2013}), in which successive atoms replace the approximate position of a neighbour --- any change in the local disclination structure must be to some extent non-local and one-dimensional-like, since it will involve a reorganization of line-defects that at each site must satisfy the SU(2) local topology.
 
The above is an example of a more general trend in which certain local geometries are more likely to facilitate LSEs. Indeed analysis of many configurations separated by 10 pico-seconds show that the topologies $(0,12,6)$, $(1,10,7)$, $(2,8,8)$, $(3,6,0)$, $(3,6,2)$, $(3,6,9)$ and ``other'' are mainly involved in LSE activity. With the exception of $(3,6,0)$ and $(3,6,2)$, these topologies correspond to regions of larger volume, which is compatible with the observation of thermally activated LSEs are more likely to occur in regions of increased free-volume~\cite{Derlet2018,Derlet2020}. The topologies of  $(3,6,0)$ and $(3,6,2)$ correspond to under-coordinated environments, and from Fig.~\ref{FigPopulation} are rare. The above result demonstrates that when they do exist, the are likely to transform into other local topologies via the thermal activation of an LSE. Finally, since thermally activated LSE activity mediates structural relaxation, it also mediates the creation of increasing $(0,12,0)$ content. The most likely initial local topologies where this transformation occurs are the $(1,10,2)$, $(2,8,2)$, $(2,8,4)$, $(3,6,4)$ and ``other'' environments. When one determines the actual number of 4- and 6-fold bonds involved in the observed transformation to the $(0,12,0)$ topology, a roughly equal number of 4- and 6-fold bonds contribute, which reflects the $n$-fold bond population gradients with respect to an increase in icosahedral content investigated in Sec.~\ref{ssec:nfold}. Fig.~\ref{FigLabelVolZ}a reveals these to have relatively low local volumes (reduced free volume content), indicating that such transformations are less likely to occur. Thus as the system relaxes, and the bond defect network evolves to a less frustrated configuration, the density of free-volume reduces, and LSEs leading to the creation of icosahedral content become less frequent resulting in slowing down of the relaxation process (as seen in Fig.~\ref{FigEvolution}).

Identifying local disclination defect structures or local regions not following the SU(2) topology rules, as regions which are more likely to undergo structural change due to thermal-activation offers the tantalizing prospect of predicting which parts of the glassy structure are likely to mediate thermally activated  relaxation and plasticity. 

It has already been established that quasi-localized low-frequency vibrational modes of model glasses have a high oscillator strength in regions that are less-likely to contain icosahedral $(0,12,0)$ and Frank-Kasper $(0,12,4)$ topologies, and are more likely to exist in regions of so-called geometrically unfavoured motives (GUMs)~\cite{Ding2014} which generally involve the Nelson SU(2) topologies. Earlier work demonstrated that such vibrational ``soft spots'' where found to correlate strongly with regions exhibiting negative  local Kelvin shear moduli which where hypothesized to more likely undergo structural rearrangement upon loading~\cite{Derlet2012}. This was confirmed in the work of Ma and co-workers~\cite{Ding2014} which found that the GUMs had a propensity to mediate plasticity under high strain rate shear loading. Later work by Falk and co-workers~\cite{Patinet2016} which exhaustively investigated what local features of the structure mediated stress-driven localized shear transformations, found little correlation with the traditional local structural indicators such as potential energy, density, the degree of short range order and local Voigt shear moduli. Whilst confirming the work of Ref.~\cite{Ding2014}, they found the strongest correlation with a local yield shear stress measure which involved shearing only cluster of atoms extracted from their simulated glassy samples. 

Refs.~\cite{Ding2014,Patinet2016} focus on predicting stress driven localized shear transformations and should be distinguished from the thermally-activated LSEs shown in Fig.~\ref{FigChange}, which under a fixed applied strain geometry also mediate thermally activated shear stress relaxation~\cite{Derlet2017a}. Indeed, it might be that predicting the location of a region susceptible to a shear stress driven instability and one prone to thermal-activation are quite different problems. The present work offers one way to study certainly the latter dynamics where the nature of the disclination network (or lack of it) forms a predictive tool to understand both thermally-activated structural relaxation and plasticity.

\section{Concluding Remarks}

A quantitative understanding of the nature of the structural constraints of a well-relaxed amorphous solid could play just as an important role as that of the theory of defects in a crystalline system. For the crystal, the reference is the long-range order of the perfect lattice, whereas for a glass the reference is most-likely a variety of low energy structural motives obeying the aforementioned constraints. These very same constraints also define the nature of the local structural excitations that will mediate glassy material evolution and its response to external stimuli as a load. The present work demonstrates that the early work of Frank, Kasper, Bernel, Turnbull and Chaudri culminating in the work of Nelson, whose mathematically rigorous description of defects in the first neighbour shell forms a consistent and possibly robust (albeit approximate) theory of bond defects within the amorphous solid. The present work finds that the vast majority of well relaxed structure have local bonding topologies that follow the prediction of the associated SU(2) algebra developed in Ref.~\cite{Nelson1983b}, both in terms of $n$-fold bond populations and also in terms of the predicted orientational geometry of the bonds. This results in a dense and extended line-defect structure of non 5-fold bonds embedded in a system spanning network of icosahedrally coordinated atomic environments. This defect network is found to exhibit a spatial icosahedral orientational correlation extending up to 3 to 4 bond-lengths. Within this context the work has numerically demonstrated that a less frustrated defect network (as defined by deviations away from equilibrium bond-lengths) will correspond to a less dense network of defect-lines and therefore an increased icosahedral content and a more relaxed glassy structure.   

The revelation that most of our glassy structure does indeed satisfy a set of local mathematically well-defined constraints, which in turn result in an emergent longer range correlation, suggests a certain level of correlated disorder. As with point, line and planar defects in crystals, the exploitation of such a disorder context, can lead to detailed theories of structural evolution involving aging and rejuvenation, and ultimately that of a thermally activated theory of plasticity. The study of atoms of different size ratios and how the addition of larger atoms affects this description forms the natural next steps to multi-compound alloy mixtures which form the industrially relevant bulk metallic glasses.

\section{Acknowledgements}

The present work was supported by the Swiss National Science Foundation under Grant No. 200021-165527. The author thanks W. L. Johnson, R. Maass and D. Rodney for useful discussions.

\end{document}